\documentclass[fleqn,usenatbib]{mnras}

\usepackage{newtxtext,newtxmath}

\usepackage[T1]{fontenc}

\DeclareRobustCommand{\VAN}[3]{#2}
\let\VANthebibliography\thebibliography
\def\thebibliography{\DeclareRobustCommand{\VAN}[3]{##3}\VANthebibliography}

\usepackage{graphicx}	
\usepackage{amsmath}	
\usepackage[dvipsnames]{xcolor}
\usepackage{pdflscape}
\usepackage{xspace}

\newcommand{\g}{\textit{Gaia}\xspace}
\newcommand{\vg}{VLBI--\textit{Gaia}\xspace}

\title[Dim cores of radio-bright AGN jets]{Dim cores of radio-bright AGN jets: VLBI and Gaia astrometry pinpoint different parsec-scale features}

\author[A. V. Popkov et al.]{
A.~V.~Popkov$^{1,2}$\thanks{E-mail: avpopk@gmail.com}, 
Y.~Y.~Kovalev$^{3}$,
A.~V.~Plavin$^{4}$,
L.~Y.~Petrov$^{5}$,
and I.~N.~Pashchenko$^{2}$
\\
$^1$Moscow Institute of Physics and Technology, Institutsky per. 9, Dolgoprudny, Moscow region, 141700, Russia\\
$^2$Lebedev Physical Institute of the Russian Academy of Sciences, Leninsky prospekt 53, 119991 Moscow, Russia\\
$^3$Max-Planck-Institut f\"ur Radioastronomie, Auf dem H\"ugel 69, Bonn D-53121, Germany\\
$^4$Black Hole Initiative, Harvard University, 20 Garden St, Cambridge, MA 02138, USA\\
$^{5}$NASA Goddard Space Flight Center, Code 61A, 8800 Greenbelt Rd, Greenbelt, 20771 MD, USA
}

\date{Accepted 2025 September 2. Received 2025 September 2; in original form 2025 June 21}

\pubyear{2025}

\begin{document}
\label{firstpage}
\pagerange{\pageref{firstpage}--\pageref{lastpage}}
\maketitle

\begin{abstract}
Astrometry with the very long baseline radio interferometry (VLBI) allows to determine the position of a point close to the source's brightest compact detail at milliarcsecond scales. 
For most active galactic nuclei (AGNs), this compact detail is the opaque core of the radio jet. 
Rare cases of sources whose brightest detail is not the core but a prominent jet feature parsecs away from the core have been reported, but such sources remained elusive. 
In this work, we use a novel method for a systematic search of these sources.
We scrutinize the AGNs for which the offset between their coordinates determined with VLBI and \g is statistically significant and coincides with the vector between two dominant features in their VLBI images, using publicly available archival multi-frequency data.
We find 35 sources whose VLBI coordinates are associated with a bright component of their jet separated by several to tens of mas from the radio core. 
Their \g coordinates, in turn, correspond to the jet origin  close to the radio jet core. 
The previously published jet directions of most of them must be reversed. 
These sources exhibit atypically low brightness temperatures of the radio cores, down to $10^9$~K in the host galaxy frame, and, at the same time, extreme brightness of the dominating jet components.
We argue that these bright components are standing shock fronts and discuss possible physical explanations for the low core brightness, such as ineffective particle heating, atypical absorption, or differential Doppler boosting.
\end{abstract}



\section{Introduction}
\label{sec:intro}

Currently, the two most precise methods to determine celestial coordinates of astronomical objects are the very long baseline interferometry (VLBI) in the radio band and the space astrometry in the optical band. Both the \g space mission \citep{2016A&A...595A...1G} and the VLBI \citep[][and references therein]{2020A&A...644A.159C, RFC} have reached a sub-milliarcsecond accuracy. For active galactic nuclei (AGNs), this corresponds to a parsec-scale resolution.

The comparison of VLBI and \g coordinates for AGNs was performed in the \g Data Release 1 \citep{2016A&A...595A...5M}, but it was focused on the agreement between positions in the VLBI and \g catalogues and concluded that, in general, the agreement was excellent. At the same time, \citet{2017MNRAS.467L..71P} focused on disagreements and showed that 6\% of the matching AGNs in the \g catalogue and the largest VLBI catalogue, the Radio Fundamental Catalogue \citep[RFC,][]{RFC}, had a \vg shift significant at the 99\% confidence level. They proved that these shifts cannot be attributed to errors of neither \g nor VLBI, and therefore manifest a genuine offset between positions at different wavelengths. Other groups made similar comparisons with smaller datasets and came to similar conclusions, using the same and subsequent releases of the \g catalogue (DR2, EDR3, DR3) and the ICRF VLBI catalogue \citep{2017ApJ...835L..30M, 2018A&A...616A..14G, 2019ApJ...873..132M, 2020A&A...644A.159C, 2022A&A...667A.148G, 2022ApJ...939L..32S}. The share of VLBI/\g matches with statistically significant offsets increased to 9\% when using the \g DR2 \citep{2019MNRAS.482.3023P} and to 11\% when using the \g DR3 \citep{2022ApJ...939L..32S} because of an improvement in the \g accuracy.

As was pointed out in \citet{2017MNRAS.471.3775P}, the \g position corresponds to the centroid of the AGN disk-jet system because \g measures the total intensity, while the VLBI position corresponds to the brightest compact feature because VLBI measures voltages that are cross-correlated. By the \vg shift we hereafter mean the offset from the VLBI position to the \g position, or, equivalently, \g coordinates minus VLBI coordinates. \citet{2017A&A...598L...1K} discovered that the \vg offsets are preferably directed either along the parsec-scale radio jet direction or opposite to it.
A detailed analysis of the \vg position offsets by \citet{2017A&A...598L...1K, 2017MNRAS.467L..71P, 2017MNRAS.471.3775P, 2019MNRAS.482.3023P, 2019ApJ...871..143P, 2020MNRAS.493L..54K} revealed a line of evidence that strongly supports the interpretation that statistically significant \vg shifts co-directional with the jet, i.~e., when the \g coordinates are further from the central engine than the VLBI coordinates, are the manifestation of bright and extended optical emission of the jets. This evidence was based on the analysis of the \vg shift directions, AGNs proper motions, colours, redshifts, optical classes, and optical polarization. According to the analysis of \citet{2019MNRAS.482.3023P}, the presence of the optical jet responsible for 50--60\% of the overall number of statistically significant \g-VLBI offsets. For the case of the \vg shifts directed opposite to the jet, these works demonstrated that the \g coordinates are associated with the accretion disk near the central supermassive black hole. Other causes of the \vg shifts were also discussed: \citet{2017ApJ...835L..30M} pointed out that the distribution of dust in galaxies may cause a \g offset with respect to the central engine; \citet{2019ApJ...873..132M} showed several examples of sources with a blended star.

To get more information from the measurements of the \vg{} shifts, it should be kept in mind that the VLBI coordinates of the AGNs are set by the position of the brightest detail of the radio jet and the dependence of its position on the frequency. In blazars, the brightest feature of the AGN structure probed by the VLBI is usually the radio core~-- the opaque apparent beginning of the jet at a given observing frequency. Synchrotron opacity causes frequency-dependent core shift, leading to offsets $\lesssim1$~mas between the true jet base and the radio core at GHz frequencies \citep{1979ApJ...232...34B, 2008AA...483..759K}. The results of the group-delay VLBI astrometry depend in this case on the geometry and other properties of the jet \citep{2009A&A...505L...1P}. For example, the group delay solution points to the true origin of the jet in the case of the conical jet geometry and the $\nu^{-1}$ dependence of the core distance on the frequency. Although recent observations and theoretical models predict deviations from both of these assumptions \citep[and references therein]{2024MNRAS.528.2523N}, we expect the synchrotron-opacity-driven \vg shifts to be at the $\sim0.1$~mas level.

However, the brightest compact radio structure detail of an AGN jet is not necessarily its core. Regions, stationary or moving, with radio brightness comparable to or higher than the core may also exist in jets parsecs away from the core. In this case, the VLBI coordinates will correspond to the position of this brightest jet detail, not the core. 
There are several well-studied examples of AGNs whose VLBI structure is dominated by a non-core feature in the jet, e.~g., 3C~119 \citep{1991A&A...245..449N}, 4C~39.25 \citep{2002ivsg.conf..233C, 2012AA...545A.113P}, or 0858$-$279 \citep{2022MNRAS.510.1480K, 2024MNRAS.528.1697K}. 
There is also a separate morphological AGN class -- compact symmetric objects; in its members, the dominating components often are the hot spots in their parsec-scale mini-lobes \citep{1994ApJ...432L..87W, 2024ApJ...961..240K}. 

Therefore, one more possible cause of a large \vg coordinate shift for an AGN is the situation when the \g coordinates correspond to the accretion disk near the central engine, while the VLBI coordinates correspond to a bright jet feature parsecs away from the radio core. In this case, the radio core is located relatively close to the \g coordinates. Several such examples have been found \citep{2021A&A...647A.189X, 2024MNRAS.528.1697K, 2025arXiv250106513F}. The fact that the VLBI and \g coordinates are associated with different separate features of the structure should be taken into account when using such sources as phase calibrators or as a part of the reference frame. Such sources are interesting not only from the astrometric, but also from the astrophysical point of view. Firstly, there should be some unusual physical conditions in them which lead to the formation of a very bright jet detail and/or to a decreased core brightness. Secondly, their jet directions determined relative to the brightest VLBI feature in a way similar to \citet{2022ApJS..260....4P} are ambiguous, because the reference point is not the core. There is also a possibility that some of such sources are double AGNs, and the details associated with both VLBI and \g coordinates are the cores of the corresponding jets, as suggested by \citet{2023ApJ...958...29C}.

There are several possible ways to search for AGNs whose compact radio structure is dominated by a non-core feature in the jet. For example, if there is a detail in the jet that is comparable or superior to the core in brightness, the VLBI astrometric solution may ``catch'' different components at different frequencies. \citet{2011AJ....142...35P} presented examples of such objects, J2020+2942/J2020+294A and J0432+4138: they have a difference at a level of over 30~mas between the position at 22~GHz and the position from dual-band 2.2/8.4~GHz observations. More such sources, colloquially called flip-floppers, were found by \citet{2013AJ....146....5P, 2022A&A...663A..83X, 2024AJ....168...76P, 2025AJ....169..173X, RFC}. Variations of the coordinates over time can also help to find them: flare activity leads to variations in the brightness of the core and may make the core the dominant structure component for a while, which causes sudden changes in the source position estimates \citep[see, e.~g.,][]{2022MNRAS.512..874T, 2024AstL...50..657O}. However, such cases are rare. At the same time, we have recently shown \citep{2021AJ....161...88P} that in a complete flux-density-limited sample selected at 1.4~GHz, the radio core does not dominate the VLBI structure in about half of the VLBI-detected sources. Therefore, non-core-dominated compact AGNs are rather numerous. An efficient way to determine where the core is in a given AGN with multiple compact components is the analysis of the spectra of the components. Since the radio core is opaque, it has a flat spectrum. In contrast, components of the optically thin radio jet have a steep spectrum.

The goal of this work is to conduct a systematic study of AGNs whose VLBI structure is dominated by a bright component of the jet, not by its core. We search for the cases where the \g coordinates correspond to the central engine, i.~e., the contribution of the optical jet is negligible, while the VLBI coordinates correspond to a prominent compact radio jet detail, brighter than the radio core.
We use the largest dataset of archival publicly available calibrated multi-frequency VLBI visibilities and images to search for objects whose \vg shift coincides with the vector between two major components of their VLBI structure. We reconstruct the spectral indices of the components, identify the actual core position, and analyse the observed characteristics of the cores and bright jet details. 

This paper is structured as follows. In \autoref{sec:select}, we describe our algorithm for searching the sources of interest and obtaining their parameters. In \autoref{sec:res}, we present our sample, examine the astrometric parameters of the found objects, and analyse the sizes and brightness temperatures of their cores and dominant jet features. We discuss the results in \autoref{sec:discus} and summarize them in \autoref{sec:sum}.

We use a convention $S\propto\nu^{\alpha}$, where $S$ is the flux density, $\nu$ is the frequency, and $\alpha$ is the spectral index. We refer to the spectrum with $\alpha<-0.5$ as steep and with $\alpha\geq-0.5$ as flat. For the size-frequency relation, we use a different sign convention, $\theta\propto\nu^{-k}$, for consistency with previous works.

\section{Sample Selection}
\label{sec:select}

\subsection{Initial search for candidates}
\label{sec:select:init}

In this work, we search for AGNs whose \g coordinates correspond to the accretion disk near the central engine, and the VLBI coordinates correspond to a bright non-core jet detail. To perform such identification, we assume that one of the two dominant details on VLBI maps of such sources is the jet component corresponding to the VLBI astrometric coordinates and the other is the actual radio core. We also assume that the angular separation between the radio core and the jet physical origin, corresponding to the \g position, is not significant compared to the size of the radio core and the measurement uncertainties. Under these assumptions, the \vg shift should coincide with the vector between two dominant components of the VLBI map.

We used optical positions from the \g Data Release 3 \citep{2023A&A...674A...1G} and VLBI positions from the Radio Fundamental Catalogue
\citep[RFC,][]{RFC}, version rfc\_2024d. 
We prefer the RFC over other VLBI astrometric catalogues because it is based on all publicly available VLBI group delays to date and has realistic estimates of position uncertainties. 
We cross-matched the two catalogues in the same way as was done by \citet{2019ApJ...871..143P}. As a result, \vg shifts were calculated for 13121 objects.

Direct cross-matching between the absolute coordinates and the VLBI map features is impossible due to the following reason. In VLBI observations, the visibility phase is distorted by atmospheric and other unaccounted delays. To deal with this, the global fringe fitting procedure is performed, which uses closure phases and a simple structure model -- usually, a delta-function in the map centre. This procedure allows to calibrate the phase and thus to make the subsequent imaging possible, but it also results in loss of the information on the absolute positions. 
Using phase referencing with respect to a nearby calibrator, one
can determine the position of the image origin with respect to the position of the VLBI calibrator. However, observations that were used for deriving the RFC were not made in the phase-reference mode.
This is why we can only deal with relative positions of VLBI map details. Therefore, we need one more assumption: what specific point of the map corresponds to the coordinates reported in the astrometric catalogues.
In recent \vg works, some authors attributed VLBI positions from an astrometric solution to the brightest features in the maps \citep{2021A&A...647A.189X, 2021A&A...651A..64L}, while others attributed VLBI positions to map centroids \citep{2024A&A...684A.202L}. In both cases, no proof was provided.
Our approach in this work is to find two dominating structure features on the VLBI map, to assume that the VLBI coordinates correspond to either of them, and then to test whether the position difference between this and another component matches the \vg offset. This and other above-mentioned assumptions limit the completeness of the resulting sample of candidates; however, they allow us to perform a uniform search through a large amount of data.

To study the radio milliarcsecond-scale structure of the sources for which we calculated \vg shifts, we used the public Astrogeo VLBI FITS image database\footnote{\url{https://doi.org/10.25966/kyy8-yp57}}.
This database contains over 120,000 VLBI maps and calibrated visibilities for the RFC sources at frequencies from 1.4 to 86~GHz, mostly at 2, 5, and 8~GHz.
To obtain the relative positions of the main structure features of these sources, we fitted brightness distribution models consisting of two circular Gaussian components to calibrated visibility data corresponding each map of each object with a calculated \vg shift. The model fitting was performed automatically with a script utilizing the \textsc{difmap} package \citep{1994BAAS...26..987S, 1997ASPC..125...77S}. The algorithm was to put the first component at the ``dirty'' map peak position, to fit this one-component model, then to put the second component at the residual map peak position and to make the second iteration of the fitting. Parameter errors were estimated following \citet{2012A&A...537A..70S}.

We then compared the \vg shift and the vector between two fitted Gaussian components. At this stage, observations of a source at different epochs and bands were treated separately. We selected the cases that satisfy all the following criteria:
\begin{enumerate}
    \item The \vg shift $r_\mathrm{vg}$ is statistically significant ($r_\mathrm{vg} > 3\sigma_\mathrm{vg}$) and large enough ($r_\mathrm{vg} > 1$~mas) to avoid the blending of the corresponding components in the VLBI maps at 2--8~GHz and to exclude the synchrotron-opacity-driven \vg shifts.
    \item The distance $r_\mathrm{comp}$ between two fitted Gaussian components of the VLBI structure is significant compared to its error $\sigma_\mathrm{comp}$ and the sizes (full widths at half maximum, FWHM) $\theta_1$ and $\theta_2$ of the components: $r_\mathrm{comp} > 3\max(\sigma_\mathrm{comp}, (\theta_1 + \theta_2)/2)$.
    \item The vectors $\boldsymbol{r}_\mathrm{vg}$ and $\boldsymbol{r}_\mathrm{comp}$ coincide with each other within the error ellipse. The axes of the error ellipse are found by adding the value of max($\sigma_\mathrm{comp}$, $(\theta_1 + \theta_2)/2$) to both of the \vg shift error ellipse axes, which, in turn, are determined using the coordinate uncertainties from the RFC and \g catalogues. 
\end{enumerate}
The condition (iii) was tested taking the centres of either of the two Gaussian components as a reference point for the VLBI coordinates. The reason is that the structure component dominating the global astrometric solution may not be the brightest one at some individual epochs and/or frequencies.

\begin{figure*}
    \centering
    \includegraphics[width=0.49\linewidth]{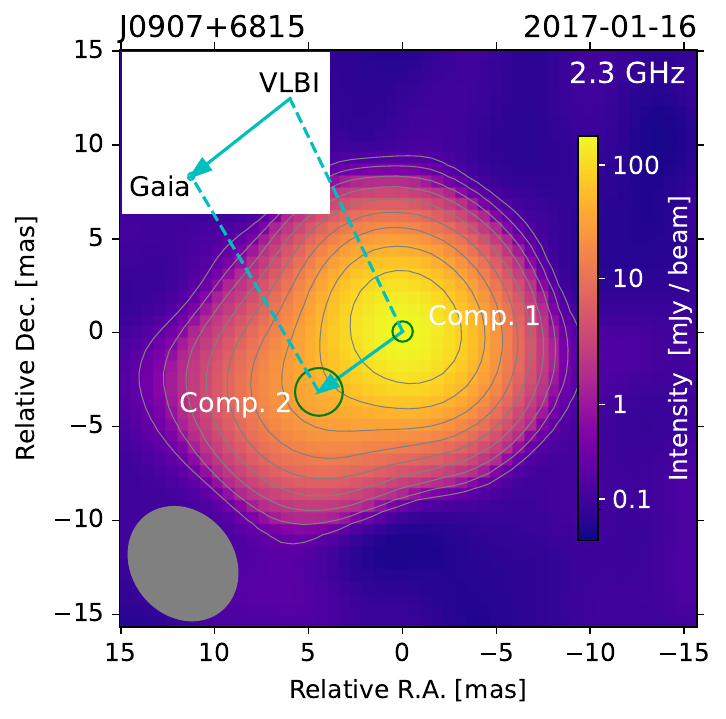}
    \includegraphics[width=0.49\linewidth]{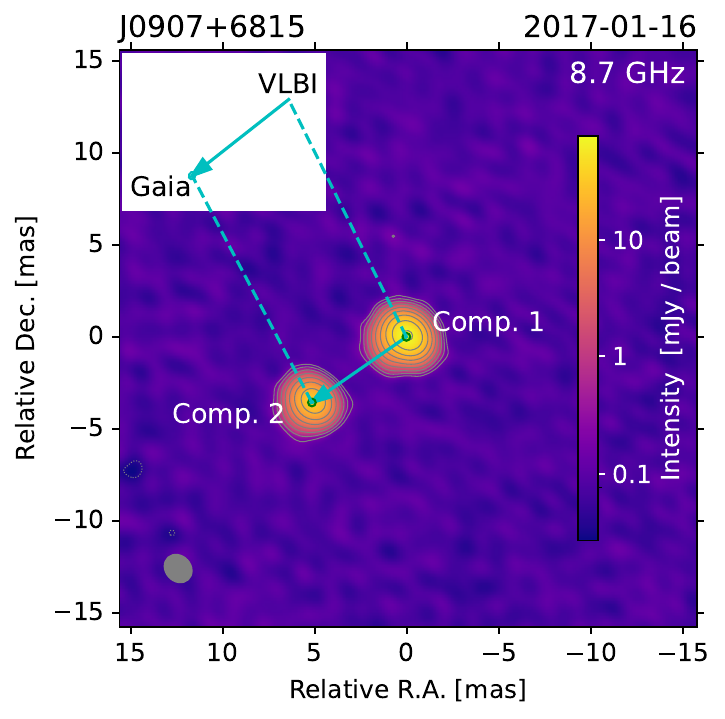}
    \caption{An example source for which we found a coincidence between the \vg shift and the vector between two dominant VLBI structure components: J0907+6815. {\it Left:} the archival VLBI map at 2.3~GHz, reconstructed by the CLEAN algorithm, from the Astrogeo database. The observing date is January~16, 2017. The pseudo-colour scale is logarithmic above the level of 10 median absolute deviations and linear below it. The contours mark the intensity starting at the level of 3 median absolute deviations with an increment factor of two. Green circles mark the positions and FWHMs of the Gaussian components approximating two dominant structure details. The absolute \vg offset vector is shown in the inset by a cyan arrow; it coincides within the errors with the vector between two major VLBI structure components, also shown in cyan. The \vg shift error ellipse is rather small; it is drawn in cyan around the arrow end. The gray ellipse in the lower left corner shows the FWHM of the CLEAN beam. {\it Right:} the same for 8.7~GHz.}
    \label{fig:maps_I}
\end{figure*}

These criteria were chosen with the aim to obtain a sample of objects in which, firstly, the coincidence between $\boldsymbol{r}_\mathrm{vg}$ and $\boldsymbol{r}_\mathrm{comp}$ is rather confident, and secondly, the determination of the components' spectral indices is possible with the existing data. The completeness of the resulting sample was not the primary goal. The limitations of this approach and the degree of completeness of the sample are discussed in \autoref{sec:discus:completeness}. We found 230 sources for which at least one observation satisfies our criteria. Hereafter we refer to these sources as candidates. \autoref{fig:maps_I} shows the examples of maps for one of them, J0907+6850, and illustrates the search procedure.

We made a rough estimate of the probability of false association (PFA) for the candidates. For each candidate, we counted the sources that meet the criteria above if the value of the \vg shift for the given candidate was used for all the sources. This number, divided by the total number of the considered sources with statistically significant shifts $>1$~mas, was adopted as the PFA. The typical obtained PFA values are of order of several percent.


\subsection{Parameters determination for candidates}
\label{sec:select:params}

The next step of our analysis was to obtain the parameters and to determine the nature of the VLBI structure components associated with the VLBI and \g coordinates, respectively. Hereafter we refer to the first of them as the VLBI-associated component, or component~1, and to the second as the \g-associated component, or component~2 (see \autoref{fig:maps_I}). 
There might be different possible combinations, of which we consider the following:
\begin{enumerate}
    \item The VLBI-associated component is the opaque radio core, and the \g-associated component is a bright detail of the jet further from the central machine than the radio core.
    \item Opposite to (i): the VLBI-associated component is a bright detail of the jet, and the \g-associated component is the opaque core.
    \item Both the VLBI- and \g-associated components are cores, i.~e., an AGN with two centres of activity is observed \citep{2006ApJ...646...49R, 2019NewAR..8601525D}.
\end{enumerate}

Opaque synchrotron-emitting regions have flat or inverted spectra, while optically thin ones have steep spectra. Therefore, to discriminate the above cases, one should inspect the spectral indices of the components. Dual- and multi-frequency VLBI data allow us to do this.

For practically all of the 230 candidates, VLBI visibilities and images at several epochs and bands are present in the Astrogeo database, with a median of five epochs per candidate. However, for most candidates, not all observations pass the selection described in \autoref{sec:select:init}. There may be physical causes of that, e.~g., a difference between the source structure at different bands or the motion of the jet components over time. However, the cause may also be artificial: if a source has a complex structure, automatic two-component modelling may not always approximate both distinct dominant structure features, even if they are present. Instead, for example, at some epochs and/or bands both Gaussians may represent the fine structure of the brightest of the two features. 

To obtain the parameters of both components of interest at the largest possible number of frequencies and epochs for each source, we made a model refinement for some observations in the following way. We visually inspected all the maps available in the Astrogeo database for all 230 candidates. If we noticed that both of the components identified by our automatic filter at one band or epoch are visible on the maps for some other observation of the source, but the brightness distribution model does not include either of them, we fitted a more complex model to the visibilities. Firstly, we tried a model consisting of three circular Gaussian components instead of two. Three-component model fitting was made automatically for all the observations of the 230 candidates in a fashion similar to the two-component modelling. For those observations of the candidates for which the three-component model also failed to approximate both components of interest, the model fitting with a sufficient number of components was made manually in \textsc{difmap}. 

Due to source variability, spectral index measurement requires simultaneous or at least close in time dual-frequency observations. For almost all candidates, there are simultaneous dual-frequency VLBI observations at 2 and 8~GHz or 5 and 8~GHz in the Astrogeo database. VLBI images at higher frequencies (15, 22, or 43~GHz) are also present for a number of sources; however, they are rarely simultaneous with 2/8~GHz and 5/8~GHz observations. For this reason, we also consider non-simultaneous observations, choosing the pairs as close in time as possible. Among these non-simultaneous pairs, 55\% are separated by less than a month and 95\% by less than half a year. In total, non-simultaneous data account for 12\% of the observations we considered for our candidates.

For the dual-frequency analysis, we need the brightness distribution models to have the components corresponding to the VLBI and \g{} coordinates at both frequencies of the VLBI observations.
With this aim in mind, for the subsequent analysis we selected only the dual-frequency pairs for which $|\boldsymbol{r}_\mathrm{vg}-\boldsymbol{r}_\mathrm{comp}| < r_\mathrm{vg} / 3$ and $\theta_1 + \theta_2 < r_\mathrm{vg}$ at both frequencies, where $r_\mathrm{comp}$ is the distance between the model components associated with the VLBI and \g{} coordinates and other designations are the same as in \autoref{sec:select:init}. 
In total, 302 dual-frequency pairs of observations for 78 candidates were selected.

\begin{figure*}
    \centering
    \includegraphics[width=\linewidth]{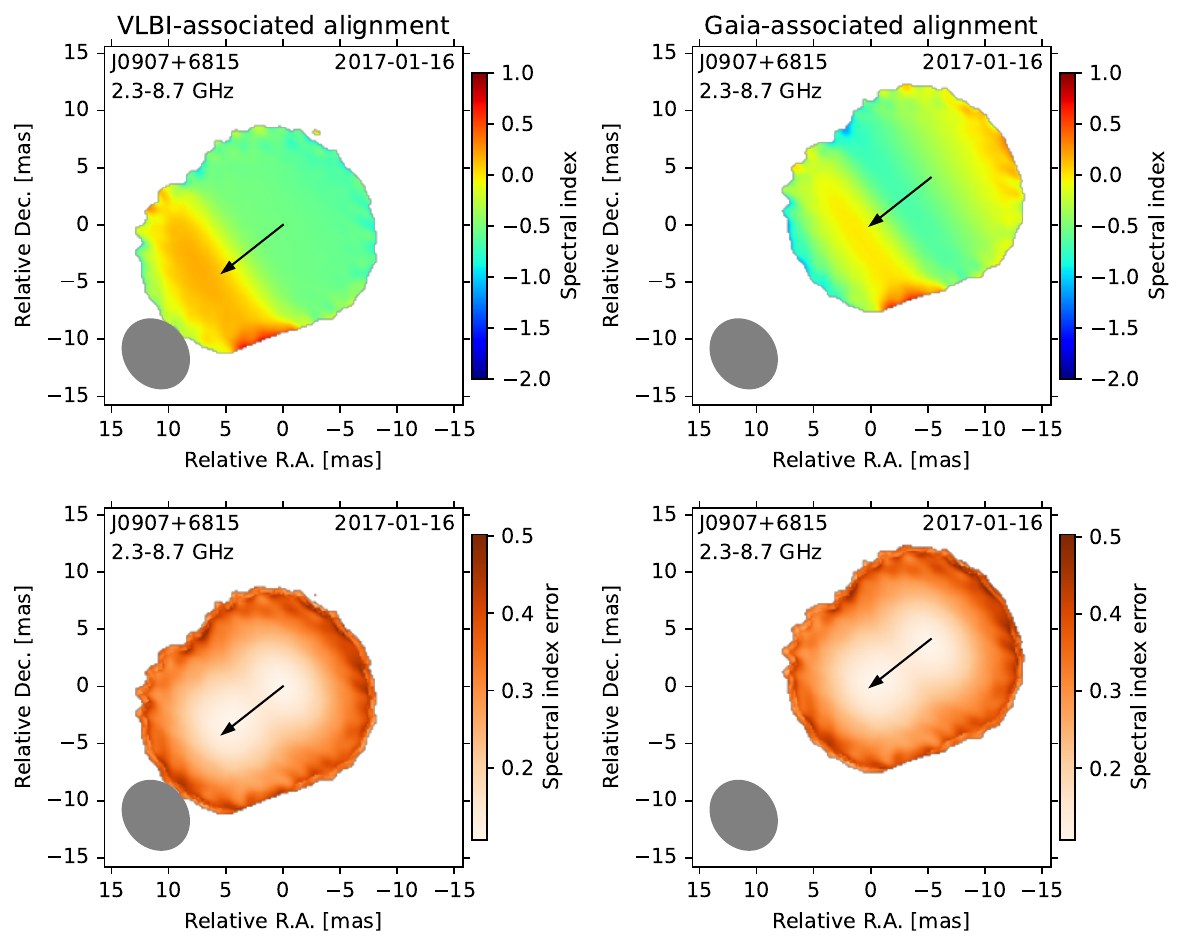}
    \caption{{\it Upper row:} spectral index maps derived from VLBI maps of the candidate source J0907+6815 at 2.3 and 8.7~GHz shown in \autoref{fig:maps_I}. {\it Lower row:} spectral index error map, calculated as described in \autoref{sec:select:params}. The beam of the map at the lower frequency (shown as the gray ellipse in the corners of all maps) was used to restore the intensity maps at both frequencies for the spectral index calculation. To align the intensity maps at the two frequencies, the positions of one of the fitted Gaussian component were superimposed: the component associated with the absolute VLBI coordinates for the left column and the component associated with the absolute \g{} coordinates for the right column. Only the alignment using the optically thin component (the VLBI-associated one in the case of this source) yields the correct results. The black arrow marks the \vg shift; its origin is aligned with the VLBI-associated component in the left column and its head is aligned with the \g{}-associated component in the right column. The region with a spectral index $\alpha > -0.5$ is the opaque jet core; for this source, it corresponds to the \g coordinates.}
    \label{fig:maps_sp}
\end{figure*}

The model parameters for the observations of this dataset are listed in \autoref{tab:single_freq}. They include the flux densities $S$, sizes (FWHM) $\theta$, and brightness temperatures $T_\mathrm{b}$ of the Gaussian components corresponding to the VLBI and \g coordinates. The brightness temperature was estimated in the observer's reference frame as:
\begin{equation}
    T_\mathrm{b} = \frac{(2\ln2)c^2}{\pi k_{B}}\frac{S}{\nu^2\theta^2}
\end{equation}
\citep[e.~g.,][]{2005AJ....130.2473K}, where $c$ is the speed of light, $k_{B}$ is the Boltzmann constant, and $\nu$ is the frequency. The errors of these parameters of the Gaussian components were estimated following \citet{2012A&A...537A..70S}. The error budget of the flux density additionally includes the VLBI amplitude calibration error, for which we adopted a relative value of 10\% \citep[see, e.~g.,][]{2012A&A...544A..34P}. For components that are unresolved given the available signal-to-noise ratio, we estimate the upper limits for the size and the lower limits for the brightness temperature following the approach of \citet{2005AJ....130.2473K}. 

For all observations in \autoref{tab:single_freq}, we ran automatic deep CLEAN imaging to obtain a more uniform set of intensity maps. The imaging procedure was an adapted version of the algorithm developed by G.~Taylor \citep{1994BAAS...26..987S}; the stopping criterion was that the residual map root mean square within the source area becomes at least three times lower than the root mean square far from the map phase centre. Such an ``overclean'' was done intentionally with the aim of minimizing the effect of the CLEAN bias, reported by \citet{2023MNRAS.523.1247P}, in the subsequent construction of spectral index maps.

The parameters derived from the dual-frequency analysis are given in \autoref{tab:pairs}. The spectral indices were obtained using two different approaches. The first one is that the spectral index $\alpha^\mathrm{gauss}$ was calculated using the flux densities of the fitted Gaussian components corresponding to the same detail at the two frequencies of the VLBI observations. We denote this spectral index of the VLBI-associated component as $\alpha_{1}^\mathrm{gauss}$ and of the \g{}-associated component as $\alpha_{2}^\mathrm{gauss}$. We also calculate and present in \autoref{tab:pairs} the power-law indices $k$ of the size-frequency relation $\theta\propto\nu^{-k}$.

\begin{landscape}

\begin{table}
\centering
\caption{Parameters of the fitted VLBI brightness distribution model components identified with the VLBI and \g coordinates. (1)~-- source J2000 name; (2)~-- observing date; (3)~-- observing frequency in GHz; (4)~-- flux density in Jy of the Gaussian component of the VLBI map corresponding to the VLBI coordinates; (5)~-- size (FWHM) in milliarcseconds of the Gaussian component corresponding to the VLBI coordinates; (6)~-- observer's frame brightness temperature in K of the Gaussian component corresponding to the VLBI coordinates; (7)-(9)~-- same as (4)-(6) for the Gaussian component of the VLBI map corresponding to the \g coordinates. The full machine-readable table is available in the online version of the article; here, first ten lines are shown for guidance.}
\label{tab:single_freq}
\begin{tabular}{ccccccccc}
\hline
Source & Date & $\nu$ & $S_{1}$ & $\theta_{1}$ & $T_\mathrm{b,1}$  & $S_{2}$ & $\theta_{2}$ & $T_\mathrm{b,2}$\\
& & GHz & mJy & mas & K & mJy & mas & K \\
(1) & (2) & (3) & (4) & (5) & (6) & (7) & (8) & (9) \\
\hline
J0000$-$3221 & 2014-06-09 & 2.3 & $628\pm72$ & $1.21\pm0.05$ & $(1.0\pm0.1)\times10^{11}$ & $65\pm15$ & $3.12\pm0.51$ & $(1.6\pm0.6)\times10^{9}$ \\
J0000$-$3221 & 2014-06-09 & 8.7 & $185\pm98$ & <1.68 & >$1.1\times10^{9}$ & $19\pm16$ & <2.67 & >$4.4\times10^{7}$ \\
J0000$-$3221 & 2017-01-16 & 2.3 & $492\pm55$ & $0.38\pm0.01$ & $(8.0\pm0.7)\times10^{11}$ & $56\pm10$ & $3.17\pm0.36$ & $(1.3\pm0.4)\times10^{9}$ \\
J0000$-$3221 & 2017-01-16 & 8.7 & $224\pm28$ & $0.33\pm0.02$ & $(3.5\pm0.5)\times10^{10}$ & $11\pm2$ & $0.57\pm0.09$ & $(5.4\pm2.0)\times10^{8}$ \\
J0001$+$0723 & 2021-10-12 & 4.4 & $61\pm8$ & $0.81\pm0.05$ & $(6.0\pm0.9)\times10^{9}$ & $12\pm3$ & $1.32\pm0.27$ & $(4.4\pm2.1)\times10^{8}$ \\
J0001$+$0723 & 2021-10-12 & 7.6 & $46\pm6$ & $0.56\pm0.04$ & $(3.1\pm0.5)\times10^{9}$ & $4\pm2$ & <0.65 & >$2.1\times10^{8}$ \\
J0008$-$2339 & 2017-05-27 & 2.3 & $393\pm98$ & $1.50\pm0.26$ & $(4.2\pm1.7)\times10^{10}$ & $94\pm31$ & <1.64 & >$8.4\times10^{9}$ \\
J0008$-$2339 & 2017-05-27 & 8.7 & $174\pm35$ & $0.44\pm0.06$ & $(1.5\pm0.5)\times10^{10}$ & $133\pm37$ & $0.42\pm0.08$ & $(1.3\pm0.6)\times10^{10}$ \\
J0008$-$2339 & 2018-06-03 & 2.3 & $405\pm110$ & $1.55\pm0.30$ & $(4.1\pm1.9)\times10^{10}$ & $63\pm28$ & <2.13 & >$3.4\times10^{9}$ \\
J0008$-$2339 & 2018-06-03 & 8.7 & $170\pm36$ & $0.46\pm0.07$ & $(1.3\pm0.4)\times10^{10}$ & $123\pm34$ & $0.46\pm0.09$ & $(9.7\pm4.6)\times10^{9}$ \\
\hline
\end{tabular}
\end{table}

\begin{table}
\centering
\caption{Parameters of the candidate sources derived from the dual-frequency VLBI data. (1)~-- source J2000 name; (2) and (3)~-- frequencies in GHz of the observations used; (4) and (5)~-- observing dates at first and second frequencies, respectively; (6)~-- spectral index of the Gaussian component of the VLBI map corresponding to the VLBI coordinates; (7)~-- the value on the spectral index map at the point corresponding to the VLBI coordinates, if the intensity maps at two frequencies are aligned using the VLBI-associated Gaussian component position; (8)~-- same as (7) with the position of the \g-associated component used for alignment; (9)~-- slope of the size vs. frequency power-law dependence for the VLBI-associated component; (10)-(13)~-- same as (6)-(9) for the \g-associated component of the VLBI map. The full machine-readable table is available in the online version of the article; here, first ten lines are shown for guidance.}
\label{tab:pairs}
\begin{tabular}{cccccrrrrrrrr} 
\hline
Source & $\nu_1$ & $\nu_2$ & Date 1 & Date 2 & $\alpha_{1}^\mathrm{gauss}$ & $\alpha_{1}^\mathrm{mapV}$ & $\alpha_{1}^\mathrm{mapG}$ & $k_{1}$ & $\alpha_{2}^\mathrm{gauss}$ & $\alpha_{2}^\mathrm{mapV}$ & $\alpha_{2}^\mathrm{mapG}$ & $k_{2}$ \\
(1) & (2) & (3) & (4) & (5) & (6) & (7) & (8) & (9) & (10) & (11) & (12) & (13) \\
\hline
J0000$-$3221 & 2.3 & 8.7 & 2014-06-09 & 2014-06-09 & $-0.91\pm0.41$ & $-1.31\pm0.11$ & $-1.13\pm0.13$ & ... & $-0.91\pm0.65$ & ... & ... & ... \\
J0000$-$3221 & 2.3 & 8.7 & 2017-01-16 & 2017-01-16 & $-0.58\pm0.12$ & $-0.62\pm0.11$ & $-0.38\pm0.11$ & $0.12\pm0.05$ & $-1.22\pm0.21$ & $-0.90\pm0.11$ & $-1.12\pm0.11$ & $1.27\pm0.14$ \\
J0001$+$0723 & 4.4 & 7.6 & 2021-10-12 & 2021-10-12 & $-0.50\pm0.34$ & $-0.56\pm0.26$ & $0.09\pm0.34$ & $0.66\pm0.17$ & $-1.85\pm0.86$ & $-3.42\pm0.55$ & $-3.91\pm0.35$ & ... \\
J0008$-$2339 & 2.3 & 8.7 & 2017-05-27 & 2017-05-27 & $-0.60\pm0.24$ & $-0.42\pm0.11$ & $-0.39\pm0.11$ & $0.92\pm0.16$ & $0.26\pm0.32$ & $-0.13\pm0.12$ & $-0.15\pm0.12$ & ... \\
J0008$-$2339 & 2.3 & 8.7 & 2018-06-03 & 2018-06-03 & $-0.65\pm0.26$ & $-0.44\pm0.11$ & $-0.40\pm0.11$ & $0.90\pm0.18$ & $0.49\pm0.39$ & $-0.11\pm0.12$ & $-0.08\pm0.12$ & ... \\
J0031$-$1326 & 4.3 & 7.6 & 2016-05-09 & 2016-05-09 & $-0.55\pm0.40$ & $-0.50\pm0.26$ & $-0.22\pm0.39$ & $1.01\pm0.23$ & $0.26\pm0.67$ & $1.18\pm0.40$ & $0.58\pm0.28$ & $-0.03\pm0.50$ \\
J0052$-$2825 & 2.3 & 8.7 & 2017-06-15 & 2017-06-15 & $-0.41\pm0.15$ & $-0.32\pm0.11$ & $-0.31\pm0.12$ & $0.81\pm0.08$ & $-0.06\pm0.29$ & $-0.25\pm0.22$ & $-0.32\pm0.20$ & ... \\
J0052$-$2825 & 2.3 & 8.7 & 2018-07-05 & 2018-07-05 & $-0.56\pm0.17$ & $-0.47\pm0.11$ & $-0.40\pm0.11$ & $0.94\pm0.10$ & $0.10\pm0.39$ & $-0.31\pm0.21$ & $-0.60\pm0.20$ & ... \\
J0112$+$4549 & 4.3 & 7.6 & 2015-12-30 & 2015-12-30 & $-0.88\pm0.51$ & $-0.74\pm0.26$ & $-0.94\pm0.33$ & $1.36\pm0.35$ & $-0.45\pm1.02$ & $-1.72\pm0.33$ & $-1.47\pm0.33$ & ... \\
J0113$+$0222 & 2.3 & 8.6 & 1999-05-10 & 1999-05-10 & $0.32\pm0.18$ & $0.28\pm0.11$ & $0.34\pm0.11$ & $0.16\pm0.11$ & $-0.61\pm0.25$ & $0.14\pm0.14$ & $0.30\pm0.15$ & $0.89\pm0.18$ \\
\hline
\end{tabular}
\end{table}

\end{landscape}

The second approach for the spectral index calculation is the construction of the spectral index maps. In order to obtain them, the images at two frequencies have to be aligned. The common practice in VLBI studies of AGN jets is to align optically thin parts of the jet to account for the core shift effect \citep[e.~g.,][and references therein]{2014AJ....147..143H}. However, for our candidates, we do not know a priori which of two dominating compact features is the core. 
For this reason, we made two versions of spectral index maps (see \autoref{fig:maps_sp} for an example). The position of the VLBI-associated Gaussian component was used for alignment in the first version, and the position of the \g-associated component -- in the second version. Of them, only the version aligned using the optically thin component provides the correct spectral index distribution. With the exception of alignment, all the other steps in producing both versions of the spectral index maps are the same. For each dual-frequency pair of observations, we restored the CLEAN maps at both frequencies using the beam of the low-frequency map. After cutting out the noise at the level of three median absolute deviations, the two-point spectral index was calculated in each map pixel. The median absolute deviation was scaled to be numerically equal to the standard deviation for the normal distribution.

The errors of the spectral index maps were calculated accounting for: (i) the statistical error, estimated from the intensity maps noise; (ii) the amplitude calibration error of 10\% of the intensity (see above); (iii) the alignment error, estimated as described below; (iv) the uncertainty of the central frequency at each band due to the combined usage of data from several frequency channels. The resulting spectral index maps may be the subject of the systematics related to the different ($u$, $v$) coverage at two bands \citep{2014AJ....147..143H} and to the CLEAN algorithm issues \citep{2023MNRAS.523.1247P}; accounting for these subtle effects is beyond the scope of this paper.
To estimate the alignment error, we generated a set of spectral index maps for slightly misaligned total intensity maps at two frequencies. The standard deviation of the spectral index within this set was considered as an estimate of the alignment error in each pixel. The artificial shifts that we introduced for this procedure lie within the circle with a radius equal to the maximum of two values: the error of the difference between the relative position of the Gaussian component, used for the alignment, on the intensity maps at two frequencies; and 10\% of the geometric mean of the major and the minor FWHM of the beam. If the number of pixels in this circular area is less than 100, we scanned all possible shifts; otherwise, 100 random shifts within this area were applied.

In \autoref{tab:pairs}, the quantities $\alpha_{1}^\mathrm{mapV}$ and $\alpha_{2}^\mathrm{mapV}$ are the values at the points associated with the VLBI and \g coordinates, respectively, on the spectral index map created using the alignment by the VLBI-associated component, while $\alpha_{1}^\mathrm{mapG}$ and $\alpha_{2}^\mathrm{mapG}$ are the similar quantities derived using the alignment by the \g-associated component. Note again that for each specific source, only one of the two alignment variants yields the correct results, depending on which of the components is optically thin.


\begin{table*}
\centering
\caption{The final sample of 35 sources for which the \g coordinates are close to the radio core and the VLBI coordinates correspond to another bright detail of the radio jet. The columns are: (1) -- source name; (2) -- source category: A -- the source belongs to the clean sample, B -- the source belongs only to the full sample (see \autoref{sec:res:sample} for details); (3) -- length of the shift from the VLBI position to the \g{} position; (4) -- \vg shift position angle (from north to east); (5) -- approximate correction to the reported by \citet{2022ApJS..260....4P} jet direction, see \autoref{sec:res:jetdir}; (6) -- median spectral index of the Gaussian component of the radio jet, associated with the VLBI coordinates; (7) -- median spectral index of the radio core component, associated with the \g{} coordinates; (8) -- median observer's frame brightness temperature of the VLBI-associated jet component; (9) -- median observer's frame brightness temperature of the radio core; (10) -- redshift from literature; (11) -- reference for the redshift. 
}
\label{tab:sample}
\addtolength{\tabcolsep}{-0.2em}
\begin{tabular}{ccrrrrrrrrl}
\hline
Source & Cat. & $\Delta r_\mathrm{VG}$ & PA$_\mathrm{VG}$ & $\Delta$PA$_\mathrm{jet}$ & $\alpha_\mathrm{jet}$ & $\alpha_\mathrm{core}$ & $T_\mathrm{b,jet}^\mathrm{obs}$ & $T_\mathrm{b,core}^\mathrm{obs}$ & $z$ & Reference \\ 
 & & mas & deg & deg & & & K & K & \\
(1) & (2) & (3) & (4) & (5) & (6) & (7) & (8) & (9) & (10) & (11) \\
\hline
J0008$-$2339 & A & $3.1\!\pm\!0.3$ & $150.6\!\pm\!2.6$ & $180$ & $-0.63\!\pm\!0.02$ & $0.38\!\pm\!0.12$ & $2.4\!\times\!10^{10}$ & $1.1\!\times\!10^{10}$ & 1.41 & \text{\citet{2004A&A...423..121S}} \\
J0031$-$1326 & A & $9.9\!\pm\!0.2$ & $77.2\!\pm\!2.4$ & $180$ & $-0.55\!\pm\!0.40$ & $0.26\!\pm\!0.67$ & $9.1\!\times\!10^{9}$ & $1.6\!\times\!10^{8}$ & ... & ... \\
J0052$-$2825 & B & $3.7\!\pm\!0.3$ & $-49.8\!\pm\!3.6$ & $180$ & $-0.49\!\pm\!0.08$ & $0.02\!\pm\!0.08$ & $3.0\!\times\!10^{10}$ & $1.5\!\times\!10^{9}$ & 1.64 & \text{\citet{2004MNRAS.349.1397C}} \\
J0154$-$0007 & A & $9.1\!\pm\!1.1$ & $141.7\!\pm\!4.4$ & $180$ & $-1.20\!\pm\!0.52$ & $-0.17\!\pm\!0.80$ & $5.5\!\times\!10^{9}$ & $1.3\!\times\!10^{9}$ & ... & ... \\
J0220$+$1652 & A & $3.8\!\pm\!0.5$ & $-50.3\!\pm\!7.5$ & $180$ & $-1.16\!\pm\!0.07$ & $0.89\!\pm\!0.32$ & $7.6\!\times\!10^{9}$ & $2.7\!\times\!10^{9}$ & ... & ... \\
J0231$-$1606 & A & $3.1\!\pm\!0.2$ & $-109.9\!\pm\!5.5$ & $180$ & $-0.83\!\pm\!0.14$ & $0.13\!\pm\!0.02$ & $1.5\!\times\!10^{10}$ & >$2.9\!\times\!10^{9}$ & 0.90 & \text{\citet{2014MNRAS.442.3329X}} \\
J0319$+$2922 & A & $2.4\!\pm\!0.4$ & $-163.0\!\pm\!9.6$ & $180$ & $-0.21\!\pm\!0.22$ & $1.45\!\pm\!0.11$ & $3.0\!\times\!10^{10}$ & $6.0\!\times\!10^{9}$ & ... & ... \\
J0402$-$1458 & B & $2.7\!\pm\!0.2$ & $-70.1\!\pm\!6.3$ & $180$ & $-0.51\!\pm\!0.14$ & $0.79\!\pm\!0.49$ & $1.7\!\times\!10^{10}$ & $1.3\!\times\!10^{8}$ & ... & ... \\
J0420$+$3849 & A & $7.3\!\pm\!0.3$ & $-109.5\!\pm\!2.0$ & $-20$ & $-0.58\!\pm\!0.17$ & $-0.30\!\pm\!0.24$ & $1.1\!\times\!10^{10}$ & $9.1\!\times\!10^{8}$ & ... & ... \\
J0523$-$2614 & A & $5.6\!\pm\!0.2$ & $116.8\!\pm\!2.4$ & $180$ & $-1.57\!\pm\!0.18$ & $-0.59\!\pm\!0.27$ & $4.1\!\times\!10^{10}$ & $7.1\!\times\!10^{9}$ & 3.11 & \text{\citet{2013AJ....146...10T}} \\
J0656$+$3209 & A & $2.7\!\pm\!0.3$ & $84.1\!\pm\!5.9$ & $180$ & $-0.35\!\pm\!0.01$ & $0.80\!\pm\!0.07$ & $1.8\!\times\!10^{10}$ & $4.4\!\times\!10^{9}$ & ... & ... \\
J0739$+$1739 & A & $1.8\!\pm\!0.4$ & $74.8\!\pm\!12.6$ & $180$ & $-0.60\!\pm\!0.21$ & $1.01\!\pm\!1.19$ & $5.9\!\times\!10^{9}$ & $1.0\!\times\!10^{9}$ & ... & ... \\
J0741$+$3112 & A & $2.3\!\pm\!0.1$ & $4.3\!\pm\!3.0$ & $0$ & $-0.74\!\pm\!0.22$ & $1.05\!\pm\!0.99$ & $9.7\!\times\!10^{10}$ & $9.6\!\times\!10^{10}$ & 0.63 & \text{\citet{2017ApJS..233...25A}} \\
J0745$+$3142 & A & $3.5\!\pm\!0.1$ & $131.6\!\pm\!2.4$ & $180$ & $-0.61\!\pm\!0.21$ & $0.71\!\pm\!0.38$ & $2.8\!\times\!10^{10}$ & $2.3\!\times\!10^{9}$ & 0.46 & \text{\citet{2017ApJS..233...25A}} \\
J0807$+$0432 & A & $5.2\!\pm\!0.2$ & $-48.5\!\pm\!2.0$ & $180$ & $-0.40\!\pm\!0.01$ & $0.42\!\pm\!0.04$ & $1.4\!\times\!10^{10}$ & $9.3\!\times\!10^{9}$ & 2.88 & \text{\citet{2017ApJS..233...25A}} \\
J0823$-$0939 & A & $17.9\!\pm\!0.3$ & $-43.5\!\pm\!0.9$ & $180$ & $-0.59\!\pm\!0.03$ & $0.43\!\pm\!0.15$ & $6.4\!\times\!10^{9}$ & $4.7\!\times\!10^{8}$ & ... & ... \\
J0907$+$6815 & A & $6.7\!\pm\!0.2$ & $128.2\!\pm\!1.7$ & $180$ & $-0.56\!\pm\!0.12$ & $-0.08\!\pm\!0.13$ & $2.7\!\times\!10^{10}$ & $2.1\!\times\!10^{9}$ & ... & ... \\
J0927$+$3902 & A & $2.8\!\pm\!0.1$ & $-80.3\!\pm\!2.3$ & $0$ & $-0.29\!\pm\!0.59$ & $0.93\!\pm\!0.41$ & $5.5\!\times\!10^{11}$ & $1.6\!\times\!10^{10}$ & 0.70 & \text{\citet{2017ApJS..233...25A}} \\
J1017$+$6116 & A & $1.6\!\pm\!0.1$ & $-101.9\!\pm\!4.7$ & $-150$ & $-0.72\!\pm\!0.41$ & $1.25\!\pm\!0.81$ & $2.1\!\times\!10^{10}$ & $1.8\!\times\!10^{10}$ & 2.77 & \text{\citet{2017ApJS..233...25A}} \\
J1146$-$2447 & A & $5.1\!\pm\!0.2$ & $167.7\!\pm\!1.4$ & $180$ & $-1.76\!\pm\!0.35$ & $-0.42\!\pm\!0.12$ & $3.2\!\times\!10^{10}$ & $5.3\!\times\!10^{9}$ & 1.94 & \text{\citet{2019MNRAS.482.3458M}} \\
J1214$+$3309 & A & $4.6\!\pm\!0.2$ & $53.9\!\pm\!1.9$ & $180$ & $-0.92\!\pm\!0.26$ & $-0.18\!\pm\!0.38$ & $2.5\!\times\!10^{10}$ & $1.7\!\times\!10^{9}$ & 1.60 & \text{\citet{2017ApJS..233...25A}} \\
J1218$+$1738 & A & $9.9\!\pm\!0.2$ & $-105.0\!\pm\!1.4$ & $180$ & $-0.48\!\pm\!0.03$ & $0.10\!\pm\!0.22$ & $2.0\!\times\!10^{10}$ & $2.0\!\times\!10^{9}$ & 1.81 & \text{\citet{2017ApJS..233...25A}} \\
J1229$+$0203 & A & $1.3\!\pm\!0.1$ & $38.2\!\pm\!5.8$ & $0$ & $-0.92\!\pm\!0.18$ & $0.90\!\pm\!0.16$ & $2.5\!\times\!10^{10}$ & $1.2\!\times\!10^{11}$ & 0.16 & \text{\citet{1992ApJS...83...29S}} \\
J1337$+$0206 & A & $6.5\!\pm\!0.2$ & $83.1\!\pm\!3.0$ & $180$ & $-0.78\!\pm\!0.07$ & $-0.42\!\pm\!0.11$ & $3.7\!\times\!10^{9}$ & $2.1\!\times\!10^{9}$ & 1.36 & \text{\citet{2017ApJS..233...25A}} \\
J1355$-$3709 & A & $9.9\!\pm\!0.6$ & $-124.5\!\pm\!3.6$ & $70$ & $-0.12\!\pm\!0.35$ & $0.87\!\pm\!0.60$ & $6.2\!\times\!10^{9}$ & $5.7\!\times\!10^{8}$ & 1.32 & \text{\citet{2022ApJ...929..108G}} \\
J1441$-$3456 & A & $8.2\!\pm\!0.5$ & $-13.3\!\pm\!1.3$ & $180$ & $-0.62\!\pm\!0.17$ & $-0.20\!\pm\!0.23$ & $1.7\!\times\!10^{10}$ & $4.2\!\times\!10^{9}$ & 1.16 & \text{\citet{2017A&A...606A..15C}} \\
J1507$+$1236 & A & $2.0\!\pm\!0.3$ & $132.3\!\pm\!8.7$ & $180$ & $-1.04\!\pm\!0.51$ & $1.13\!\pm\!0.95$ & $6.9\!\times\!10^{9}$ & >$2.9\!\times\!10^{9}$ & 1.54 & \text{\citet{2009ApJS..180...67R}} \\
J1602$+$2418 & A & $5.6\!\pm\!0.2$ & $-65.4\!\pm\!2.1$ & $180$ & $-0.75\!\pm\!0.04$ & $-0.31\!\pm\!0.06$ & $5.2\!\times\!10^{9}$ & $6.5\!\times\!10^{9}$ & 1.79 & \text{\citet{2017ApJS..233...25A}} \\
J1630$-$1157 & A & $42.1\!\pm\!0.4$ & $-110.5\!\pm\!0.7$ & $-140$ & $-0.46\!\pm\!0.19$ & $-0.15\!\pm\!0.20$ & $2.1\!\times\!10^{9}$ & $3.3\!\times\!10^{8}$ & ... & ... \\
J1641$-$1754 & A & $13.2\!\pm\!2.8$ & $31.5\!\pm\!8.1$ & $...$ & $-0.78\!\pm\!0.48$ & $0.04\!\pm\!0.69$ & $2.1\!\times\!10^{9}$ & >$7.0\!\times\!10^{8}$ & ... & ... \\
J1754$+$6452 & B & $2.7\!\pm\!0.2$ & $-72.4\!\pm\!5.1$ & $160$ & $-0.68\!\pm\!0.02$ & $0.68\!\pm\!0.16$ & $2.0\!\times\!10^{10}$ & $8.2\!\times\!10^{8}$ & 0.98 & \text{\citet{2024AJ....168...58D}} \\
J1756$+$5748 & A & $9.4\!\pm\!0.2$ & $-101.4\!\pm\!1.1$ & $180$ & $-1.17\!\pm\!0.10$ & $-0.29\!\pm\!0.01$ & $6.9\!\times\!10^{9}$ & $1.2\!\times\!10^{9}$ & 2.11 & \text{\citet{1997MNRAS.290..380H}} \\
J1821$+$3945 & A & $28.6\!\pm\!0.2$ & $-148.6\!\pm\!0.5$ & $-130$ & $-0.64\!\pm\!0.13$ & $-0.14\!\pm\!0.04$ & $3.6\!\times\!10^{9}$ & $7.0\!\times\!10^{8}$ & ... & ... \\
J2149$+$0756 & A & $8.7\!\pm\!0.3$ & $15.4\!\pm\!1.8$ & $180$ & $-0.74\!\pm\!0.12$ & $-0.15\!\pm\!0.13$ & $1.1\!\times\!10^{10}$ & $1.4\!\times\!10^{9}$ & 0.52 & \text{\citet{2019ApJS..240....6Y}} \\
J2344$+$2952 & B & $3.2\!\pm\!0.4$ & $-80.7\!\pm\!6.7$ & $180$ & $0.05\!\pm\!0.18$ & $0.47\!\pm\!0.41$ & $3.6\!\times\!10^{9}$ & $6.6\!\times\!10^{8}$ & 0.95 & \text{\citet{2013AJ....146...10T}} \\
\hline
\end{tabular}
\addtolength{\tabcolsep}{0.2em}
\end{table*}

\section{Results}
\label{sec:res}

In \autoref{sec:select:params}, we listed three most probable alternatives for the nature of two VLBI structure components of an AGN jet, when one of them corresponds to the VLBI absolute position, while another~-- to the \g absolute position. We also have derived the parameters of the VLBI- and \g-associated components. In this section, we use this information to select the final sample of objects and analyse their properties.

The 78 candidates, for which we measured the VLBI structure parameters of the VLBI- and \g-associated components at two or more frequencies, separate into two groups. Within the first group, the VLBI-associated component is optically thick, having a flat spectrum, while the component of the VLBI map associated with the \g coordinates is optically thin and has a steep spectrum. The second group represents the opposite case: the steep-spectrum VLBI-associated component and the flat-spectrum \g-associated component. In the first case, as was discussed in \autoref{sec:intro}, the VLBI coordinates point to the core region, and the \vg{} shift is likely explained by a significant contribution of the jet emission in the optical band. In this case, the interpretation of a coincidence between the optical centroid of the disk-jet system and the radio jet component distant from the core requires in-depth analysis that is beyond the scope of this paper. At the same time, the second case, when the VLBI astrometry points to a non-core feature of the jet, while the \g coordinates are close to the radio core position, is exactly the case of our interest. It is related to an unusual parsec-scale radio morphology and has been less studied so far. Hereafter, we focus on the second case. We did not find any reliable candidates for the third case~-- two optically thick cores of an AGN with two activity centres, like the one observed by \citet{2006ApJ...646...49R}. This is in agreement with the results of \cite{2011MNRAS.410.2113B}, who found only one previously known binary supermassive black hole candidate among 3114 radio-loud AGNs using a VLBI dataset partially overlapping with ours.

\subsection{Final sample}
\label{sec:res:sample}

To our final sample we select the candidates in which the VLBI-associated component is a non-core feature in the radio jet and the \g-associated component is the opaque jet core. Therefore, we require that, at least at one epoch, $\alpha_{2} \ge -0.5$, where $\alpha_\mathrm{2}$ is the spectral index of the \g-associated component, and the VLBI-associated component has a steeper spectrum than the \g-associated one. In a more accurate form, these conditions can be formulated as follows:
\[\begin{cases}
    \alpha_{2} + \sigma_{\alpha,2} \ge -0.5 \,,\\ 
    \Delta\alpha \equiv \alpha_{2} - \alpha_{1} > \sigma_{\Delta\alpha} \,,
\end{cases}\]
where $\alpha_\mathrm{1}$ and $\alpha_\mathrm{1}$ are the spectral indices of the VLBI- and \g-associated components, correspondingly, and $\sigma_{\alpha,2}$ and $\sigma_{\Delta\alpha}$ are the errors of $\alpha_{2}$ and $\Delta\alpha$. For this particular comparison, we do not account for the amplitude calibration uncertainty in $\sigma_{\Delta\alpha}$, since calibration issues affect the VLBI- and \g-associated components similarly. The conditions above were complemented by the criterion that the VLBI-associated component is brighter than the \g-associated component, $T_\mathrm{b,1} > T_\mathrm{b,2}$, at least at one frequency. Otherwise, the identification of the corresponding Gaussian component with the VLBI coordinates reference point, made at the previous stages of our analysis, would be highly likely spurious.

As described in \autoref{sec:select:params}, we calculated the spectral indices of the components by two methods: using the flux densities of the fitted Gaussian components and using the spectral index maps. For the spectral index maps, we tried two variants of the alignment: by the position of the VLBI-associated component and by the position of the \g-associated component. Since at this final stage we select the sources for which \g coordinates are associated with the core, the correct variant in this case is the alignment by the position of the VLBI-associated component. 

\begin{figure}
    \centering
    \includegraphics[width=\columnwidth]{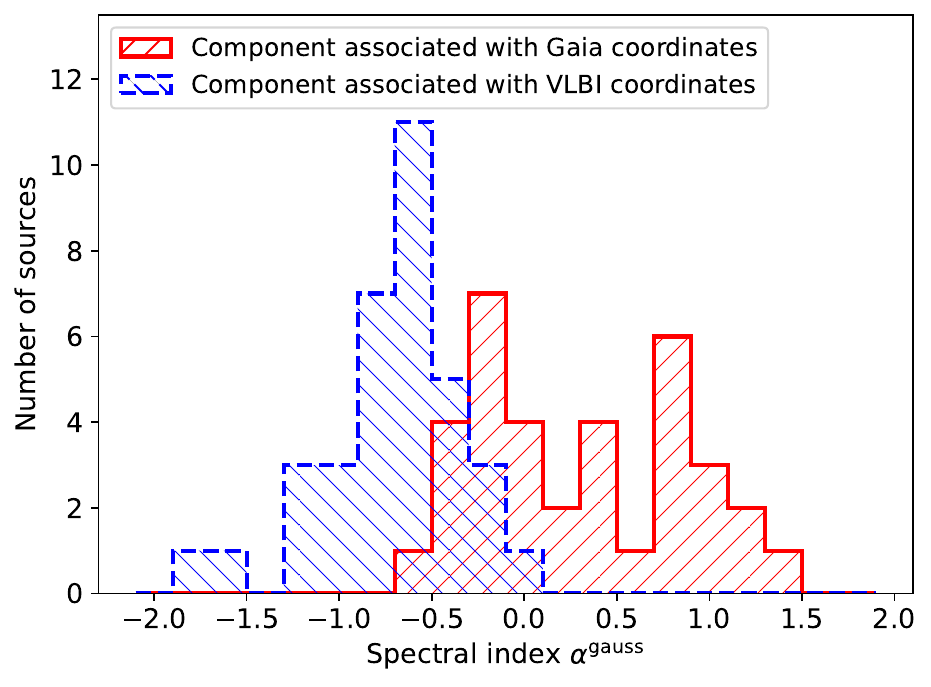}
    \caption{Distributions of the median VLBI spectral indices of the components for 35 sources selected to the final sample. The red histogram represents the radio core components associated with the \g coordinates; they have flat or inverted spectra. The blue histogram represents the jet components associated with the VLBI coordinates; most of them are optically thin and each of them has a steeper spectrum than the corresponding \g-associated component.}
    \label{fig:sp_ind}
\end{figure}

If we consider only the spectral indices from the first method ($\alpha^\mathrm{gauss}$), the selection results in the sample of 35 sources. For 31 of them, the spectral indices determined by the second method ($\alpha^\mathrm{mapV}$) confirm this identification. We consider these 31 objects to be the most confident candidates, forming the sample we hereafter refer to as ``clean''. However, we include all the 35 sources in our final sample (hereafter ``full''). They are listed in \autoref{tab:sample}, where the sources from the clean sample are marked as category A, and other sources -- as category B.  \autoref{tab:sample} contains the parameters we determined for the sources and the redshifts collected from the literature using the NASA Extragalactic Database (NED)\footnote{\url{https://ned.ipac.caltech.edu/}}. For the spectral indices and brightness temperatures, the median values are given. The median spectral indices of the Gaussian components were computed across epochs and frequency pairs that met the above conditions. The distributions of the median spectral indices are shown in \autoref{fig:sp_ind}. The median brightness temperatures were calculated in logarithm using all observations of each source where the VLBI–Gaia offset matched the component separation vector in two or more frequency bands (all observations from \autoref{tab:single_freq}, see \autoref{sec:select:params}).
There are two prominent sources in our sample: 4C~39.25 (J0927+3902), previously known to have a non-core-dominated VLBI structure \citep{2002ivsg.conf..233C, 2012AA...545A.113P}, and the prototypical quasar 3C~273 (J1229+0203). For the latter, the core is generally the brightest component of the VLBI structure; however, at several observing epochs, another jet component became the brightest, and these epochs were selected by our procedure. The fact that there are stationary features in the innermost jet of 3C~273 with a brightness comparable to the core was known before \citep{2017MNRAS.468.4478L}. These bright components may cause the shift of the astrometric VLBI position of 3C~273 from its radio core \citep{2017MNRAS.471.3775P}.

\subsection{Astrometry of the final sample}
\label{sec:res:astrometry}

\begin{figure}
    \centering
    \includegraphics[width=\columnwidth]{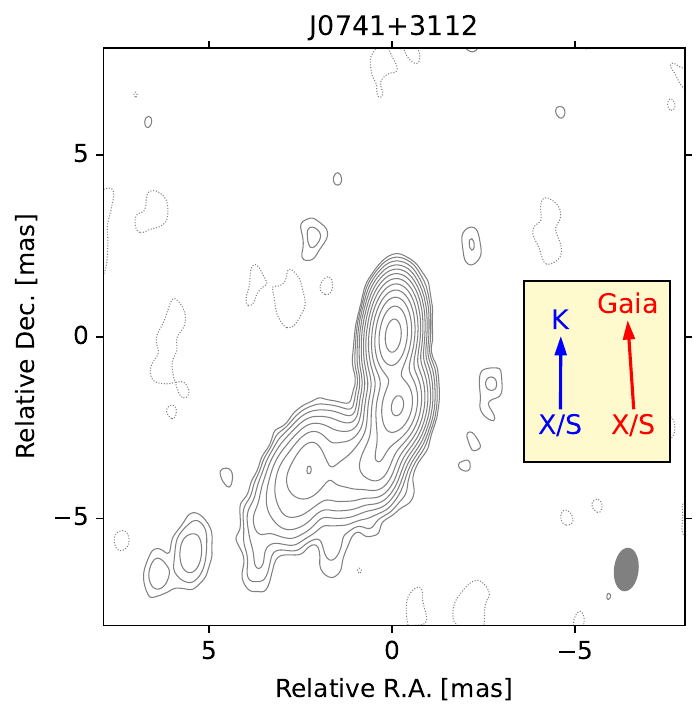}
    \caption{An example illustrating pitfalls of multi-band astrometry for the AGNs with bright jet components: a 15~GHz VLBI contour map from the MOJAVE project archive \citep{2018ApJS..234...12L} for the object J0741+3112 with a significant difference between positions at different bands. The offsets between the absolute coordinates from the RFC dual-band X/S (8/2~GHz) solution and the RFC K-band (24~GHz) and \g{} solutions, correspondingly, are shown by the arrows in the inset. The X/S position is associated with the bright jet component dominating at lower frequencies, while the K-band and the optical positions -- with the core region. See \autoref{sec:res:astrometry} for details.}
    \label{fig:mwl_coord}
\end{figure}

Our final sample consists of the sources for which VLBI astrometry determines the coordinates of a jet component that is not the core. In this subsection, we analyse how this fact affects the astrometric results. 

We should note that most of the 35 AGNs of our sample are well known astrometric sources, listed in VCS1 \citep{r:vcs1}, ICRF2 \citep{r:icrf2}, ICRF3 \citep{2020A&A...644A.159C}, WFCS \citep{r:wfcs}, VIE2022b \citet{r:kra23}, OBRS2 \citep{2013AJ....146....5P}, VIPS \citep {r:astro_vips}, and other catalogues. They were also observed in regular geodetic observations. When two parameters of source positions are estimated from a global dataset, the uncertainty of the source position depends on the signal-to-noise ratio, recorded bandwidth, the number of observations, and the geometry of the network. Therefore, peculiarities of these sources cannot be noticed from analysis of their position uncertainties. This is verified by the comparison we made for the distributions of the right ascension and declination errors between our sample and the entire RFC. At the same time, the presence of a second compact component causes significant source structure delay, much greater than for sources having a compact core and a diffuse jet. Since the source structure is not yet accounted for in routine data analysis, the positions of these sources suffer from greater than average systematic errors that could manifest themselves in a larger scatter in position time series and larger post-fit residuals.

We also checked the absolute astrometric proper motions of the sources from our sample. Of course, intrinsic proper motions of AGNs are infinitesimal because they are at gigaparsec distances. However, AGN variability causes both jitter of their optical centroids and changes in their radio jet structure, which leads to apparent proper motions in the \g and VLBI solutions, respectively. For example, some sources with significant \vg offsets exhibit significant proper motions parallel to the jets, as demonstrated by \citet{2019MNRAS.482.3023P}. We compared the proper motions in our sample with those in a control sample of 200 random sources from the RFC for which the VLBI observations cover at least two years. In terms of the \g DR3 proper motions, there is no significant difference between our sample and the control sample. Both samples have median proper motion values approximately at the noise level. This is in agreement with our interpretation that, in the sources of our sample, the \g coordinates point to the central engine for which no significant proper motion is predicted.

To analyse the VLBI proper motions, we ran a RFC-type solution with an expanded list of sources whose proper motions were estimated, including our candidates and the sources of the control sample described above. Since for AGNs in our sample, a non-core feature, which may be non-stationary, dominates the VLBI astrometric solution, these sources may exhibit considerable VLBI proper motions. However, we found no statistically significant differences in the VLBI proper motions or their errors between the target and control sample, regardless of whether a full or clean sample was used as the target one. This may indicate that the dominant VLBI features are mostly quasi-stationary in our sample. However, the size of our sample is not enough to draw a confident conclusion. Ten sources have proper motions at the $>3\sigma$ level; of them, the largest values are $0.22\pm0.06$~mas~yr$^{-1}$ for J0739+1739, $0.18\pm0.03$~mas~yr$^{-1}$ for J1756+5748, $0.12\pm0.03$~mas~yr$^{-1}$ for J0523$-$2614, and $0.11\pm0.01$~mas~yr$^{-1}$ for J0807+0432. Note that in our sample, the prevalence of sources with a stationary jet component may be a result of the selection bias: the VLBI imaging observations we analysed are not simultaneous with the mean epoch of neither the RFC nor the \g{} catalogue, so the coincidence between the \vg{} shift and the separation of the structure components is more likely if the components are stationary.

As discussed in \autoref{sec:intro}, another characteristic effect for sources like ours is a significant difference in the VLBI coordinates at different bands. Throughout the paper, we use the results of the RFC fused solution that combines observations at all available bands as described in \citet{RFC} as the VLBI coordinates. RFC also contains the multi-band extension, providing the results of the dual-band (2~GHz~/~8~GHz or 5~GHz~/~8~GHz) and different single-band solutions. We inspected the difference between the positions from the dual-band solution and from the solution at the highest of the provided frequencies, 24~GHz (K band), for our sample. We expect that the core dominance increases with frequency, since the core has a flat spectrum, while other regions of the jet have a steep spectrum. The K-band coordinates are available for only 5 of the 35 objects in our sample. For one of them, J0741+3112, the discussed effect is indeed present. \autoref{fig:mwl_coord} shows a 15~GHz VLBI map of this source from the MOJAVE project archive\footnote{\url{https://www.cv.nrao.edu/MOJAVE/}}; no 24~GHz maps are present in the Astrogeo database for this object. 
The structure of this source is dominated by two compact components of comparable brightness, separated by about 2~mas. The southern one is a quasi-stationary detail in the jet; we associate it with the VLBI coordinates from both the dual-band and the fused solutions. The northern one is the core; both the VLBI K-band and the \g positions are close to it. The offset from the dual-band to the K-band VLBI positions is $(-0.01\pm0.13)$~mas in the right ascension and $(1.83\pm0.13)$~mas in the declination. The value of the \vg shift is $(0.15\pm0.12)$~mas in RA and $(2.26\pm0.13)$~mas in Dec. Both of these offsets are shown in the inset of \autoref{fig:mwl_coord}.
One can see that the discussed positional difference cannot be explained by the measurement errors. The physical explanation is much more plausible: at low frequencies, the jet component dominates; at the K-band and higher frequencies, the core becomes the brightest detail. In fact, it becomes the brightest even at 15~GHz, as the map shows. 


\subsection{Jet directions}
\label{sec:res:jetdir}

\citet{2022ApJS..260....4P} published the automatically determined jet directions at the VLBI scales for thousands of AGNs. The direction was measured with respect to the brightest component. Usually, the brightest component of the VLBI structure of an AGN is the jet radio core. For our sample, this is not the case. Therefore, for most of the sources in our sample, a correction of the previously published jet direction is needed, taking into account the updated identification of the core. We specify these approximate corrections in \autoref{tab:sample}. For most sources, the jet direction must simply be reversed, i.~e., the correction is $180\degr$. For some sources with a curved jet, the correction for the innermost jet direction is different. For several sources, the correction is zero, which means that the core was correctly identified by \citet{2022ApJS..260....4P}. We discuss the impact of this correction of the
jet direction in \autoref{sec:discus:completeness}.


\subsection{Size-frequency dependence}

\begin{figure}
    \centering
    \includegraphics[width=\columnwidth]{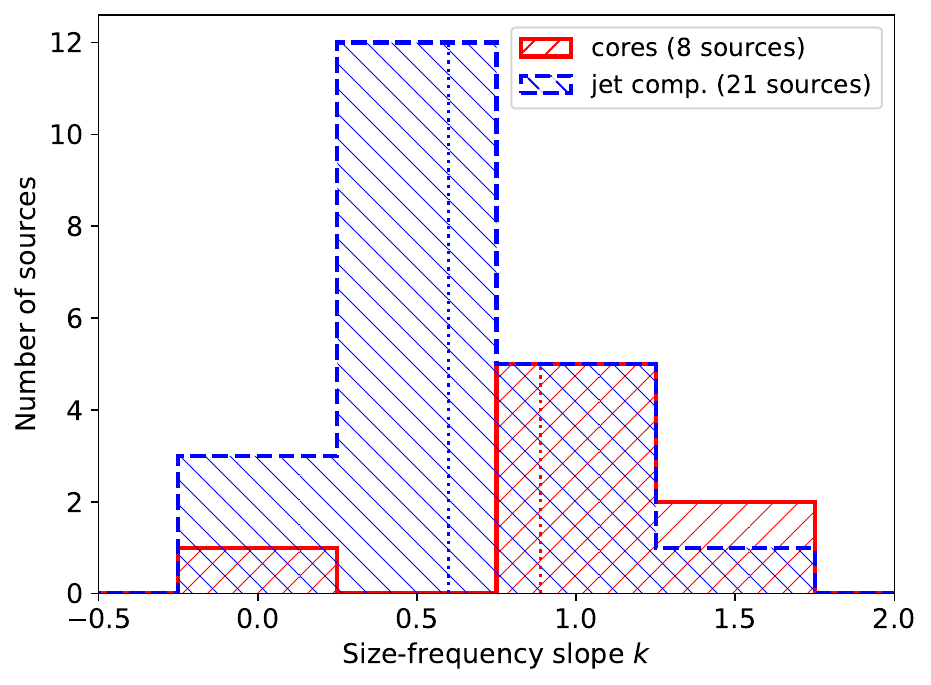}
    \caption{Distributions of the median power-law index of the size-frequency relation ($\theta\propto\nu^{-k}$) within our clean sample. The histograms for both the jet components corresponding to the VLBI coordinates and opaque cores corresponding to the \g coordinates are shown. Only the measurements with an error of $k$ less than $0.15$ were accounted. The sample median values are marked by the vertical dotted lines. The distribution for the cores has a median close to $k=1$, in agreement with previous studies. The peak for jet components at $k\approx0.5$ indicates that they likely are shock fronts.}
    \label{fig:k_size}
\end{figure}

It was shown both theoretically and observationally \citep{1979ApJ...232...34B, 2015MNRAS.452.4274P, 2022MNRAS.515.1736K} that if interstellar scattering is negligible, the AGN radio core size scales with frequency as $\theta\propto\nu^{-k}$, where $k\approx1$. For optically thin jet components, $k$ depends on their nature. In general, one expects no size-frequency dependence if the components are transparent plasma blobs. However, if the components are shock fronts or reconnection layers with the ongoing particle acceleration, then the highest frequency emission originates near the shock front while the low frequency emission originates from larger volumes behind the shock \citep{1985ApJ...298..114M}. In such cases, the predicted frequency dependence of the layer thickness of the emitting particles in case of synchrotron cooling is $\sim \nu^{-0.5}$ \citep{2013MNRAS.436.3341Z}. 

We tested these relations for our sample, considering both the jet components corresponding to the VLBI coordinates and the opaque jet cores corresponding to the \g coordinates. The values of $k$ are given in \autoref{tab:pairs}. The distributions of the $k$ index for both components are shown in \autoref{fig:k_size}; for each source, the median value of $k$ is used for each component. Only the sources of the clean sample and only the measurements with a $k$ error $<0.15$ are included in the distributions in \autoref{fig:k_size} and in the analysis in this subsection. The results for the full sample are generally the same, but with a lower significance.

The characteristic values of the $k$ index for the dominating jet components and for the cores identified by our spectral index analysis are different. The median $k$ is $0.89\pm0.16$ for the cores and $0.60\pm0.05$ for the VLBI-associated components (errors are estimated using the bootstrapping). 
The one-sided permutation test gives the $p$-value of 3\% for the difference of the median $k$ for the cores and the jet components.
This difference in the $k$ index between the cores and the jet components is theoretically predicted, as discussed above.
The median value obtained for the cores, $k\approx1$, is in agreement with previous studies. 
Regarding the VLBI-associated components of the jets, it is interesting that their $k$ distribution peaks around $k\approx 0.5$. This could imply that VLBI-associated components are shock fronts or reconnection layers with the ongoing particle acceleration. Given the absence of excess proper motions for these details in our sample reported in \autoref{sec:res:astrometry}, we suggest that the dominating jet details are stationary shocks at least in a significant fraction of the sources of our sample. Previous studies have found quasi-stationary features in a number of AGN jets \citep[e.~g.,][]{2021ApJ...923...30L}; the speciality of our sources is that such features dominate their VLBI structure. Note, however, that the circular Gaussian model we use for the components may not adequately describe a post-shock emitting region, which may cause a bias in our measurements of the component sizes.


\subsection{Low brightness of the cores and enhanced brightness of jet details}
\label{sec:res:tb}

\begin{figure}
    \centering
    \includegraphics[width=\columnwidth]{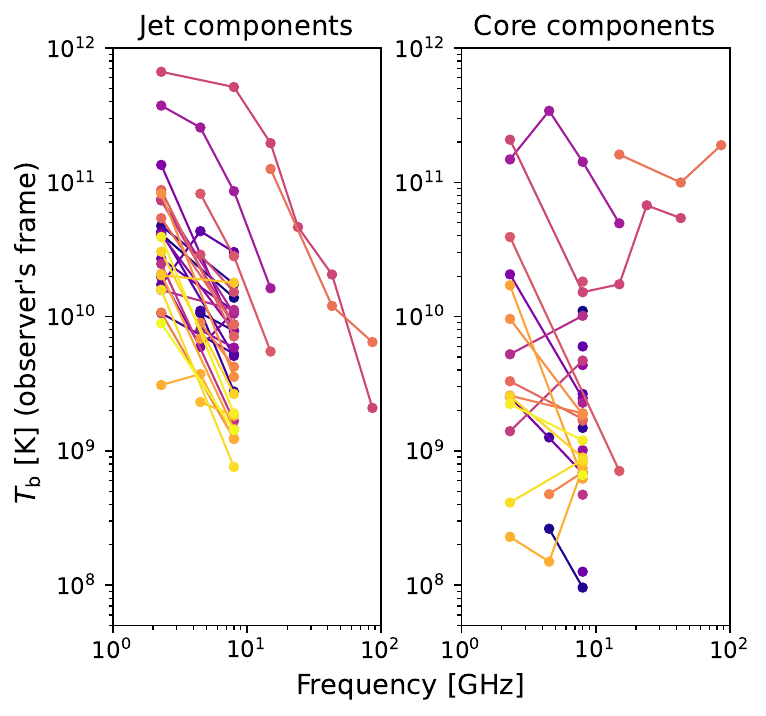}
    \caption{Frequency dependence of the median brightness temperatures in the observer's reference frame for the jet components corresponding to the VLBI coordinates (left) and opaque cores corresponding to the \g coordinates (right) in our sample. The measurements for different sources are shown in different colours; the colours are the same for both panels. The core $T_\mathrm{b}$ generally has lower values and flatter frequency dependence than the $T_\mathrm{b}$ of jet components.}
    \label{fig:tb_spectra}
\end{figure}

\autoref{tab:sample} contains the median brightness temperatures of the radio jet components, associated with the VLBI coordinates of the sources in our final sample, and of their radio cores, associated with the \g coordinates. One can see that the VLBI-associated jet component typically has several times higher $T_\mathrm{b}$ than the core and that the cores have unusually low brightness in our sample. This is exactly the reason why the VLBI astrometry determines the coordinates of the non-core detail in these sources. Only two sources in our sample have the median core $T_\mathrm{b}$ higher than the median $T_\mathrm{b}$ of the jet component: J1229+0203 and J1602+2418. However, for J1602+2418, the values are comparable; for J1229+0203, the core is generally the brightest component of the VLBI structure, except for a few epochs, as noted in \autoref{sec:res:sample}.

\autoref{fig:tb_spectra} illustrates the difference between the cores and the jet components in terms of brightness temperature. For all sources of our sample, the observer's frame $T_\mathrm{b}$ of both components is shown as a function of the observing frequency. Each $T_\mathrm{b}$ value in this plot is a median across all observing epochs from \autoref{tab:single_freq} at the given frequency. The error bars are not shown because the $T_\mathrm{b}$ values are model-dependent and we estimate that they are accurate within a factor of two. The brightness temperature of most jet components steeply decreases with frequency, as expected for optically thin details. The core brightness temperatures of individual sources have various frequency dependences, but in the sample in general, the trend of the core brightness temperature with frequency is weak. This is in agreement with theoretical predictions \citep{1979ApJ...232...34B} and with dedicated observational multi-frequency studies of the core brightness temperature \citep{2014JKAS...47..303L, 2020ApJS..247...57C, 2025A&A...695A.233R}. The weak dependence of the core brightness temperature on the frequency allows us to use the median core $T_\mathrm{b}$ across all observing frequencies in the comparison below.

\begin{figure}
    \centering
    \includegraphics[width=\columnwidth]{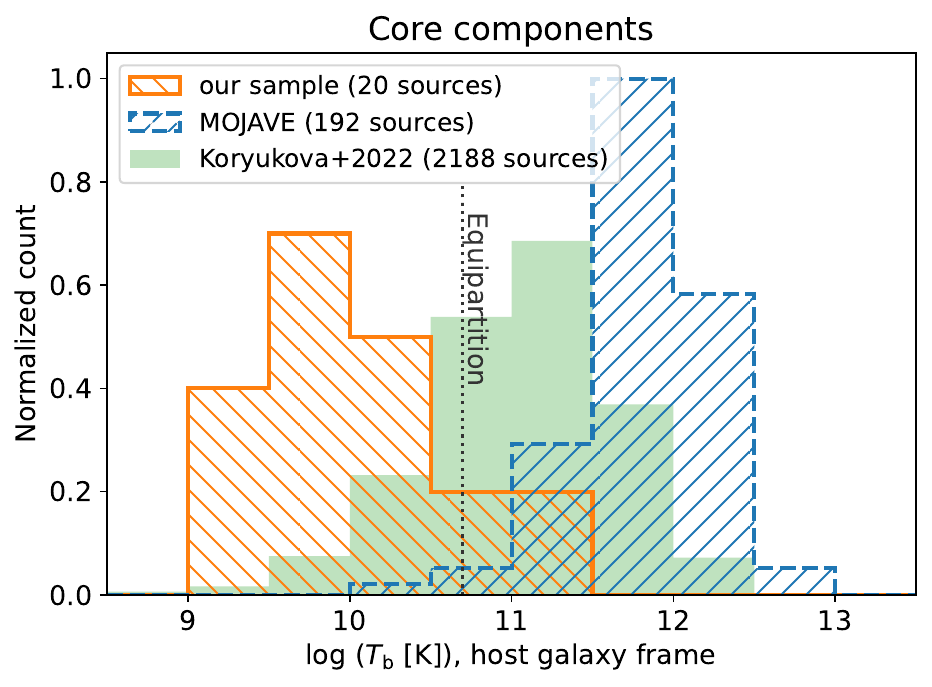}
    \caption{Empirical probability density histograms for the median of the logarithm of the radio core brightness temperature for our sample (20 sources with known redshifts and measured core $T_\mathrm{b}$). Two flux-density-limited AGN samples are shown for comparison: the MOJAVE 1.5~Jy Quarter Century sample \citep[192 sources with known redshifts and measured core $T_\mathrm{b}$,][]{2021ApJ...923...67H} and a larger sample of RFC sources with a 8~GHz flux density >0.15~Jy from the \citet{2022MNRAS.515.1736K} dataset. The core $T_\mathrm{b}$ values are given in the host galaxy reference frame (redshift correction applied, Doppler factor correction not applied). The vertical dotted line marks the equipartition $T_\mathrm{b}$ estimate by \citet{1994ApJ...426...51R}. In our sample, the radio cores of most sources have atypically low brightness temperatures.}
    \label{fig:tb_core}
\end{figure}

For the sources with available redshift measurements, we plot in \autoref{fig:tb_core} the distribution of the median core brightness temperatures in the reference frame of the host galaxy. Hereafter in this subsection, we discuss our full sample that includes the sources of both the categories A and B from \autoref{tab:sample}. All the qualitative results reported below are valid also for the clean sample (category A only). The distribution of the core brightness temperatures for our sample has a peak at $T_\mathrm{b}\sim10^{9.5}-10^{10}$~K. This is atypically low for radio cores of relativistic jets in AGNs. The classical estimate under the assumption of the energy equipartition between the radiating particles and the magnetic field by \citet{1994ApJ...426...51R} yields $T_\mathrm{b}\approx5\times10^{10}$~K in the jet plasma's rest frame. $T_\mathrm{b}$ in the host galaxy frame is expected to be about an order of magnitude higher due to the Doppler boosting. Indeed, the analysis of the long-term monitoring of the AGN brightness temperatures and kinematics by the MOJAVE team showed that the median intrinsic $T_\mathrm{b}$ of the cores is $(4.1\pm0.6)\times10^{10}$~K, in agreement with the equipartition estimate,
and typical Doppler factors are $\sim\!10$ \citep{2021ApJ...923...67H}. 
In \autoref{fig:tb_core}, we show for comparison the median core $T_\mathrm{b}$ distributions in two complete flux-density-limited AGN samples. One of them is the MOJAVE 1.5~Jy Quarter Century sample \citep{2021ApJ...923...67H}. Another is the sample of RFC sources with a 8~GHz flux density >0.15~Jy, for which we used the core parameters determined by \citet{2022MNRAS.515.1736K}. The RFC 0.15~Jy sample includes the MOJAVE 1.5~Jy sample. However, the distribution for the former is shifted to lower values due to several factors. Firstly, the RFC 0.15~Jy sample is about ten times more numerous than the MOJAVE 1.5~Jy sample and includes, on average, weaker sources than MOJAVE. Secondly, all the MOJAVE data were obtained at 15~GHz, while for the RFC 0.15~Jy sample sources, we calculated the median across the measurements at all frequencies, mostly lower than 15~GHz. Thirdly, different modelling techniques were applied in the two datasets: in MOJAVE, the core was approximated by an elliptical Gaussian, and the jet -- by the CLEAN components, while \citet{2022MNRAS.515.1736K} simply fitted the whole source structure by two circular Gaussians. The latter approach is more similar to that used in the current work.

The typical core brightness temperatures in our sample are about two orders of magnitude lower than those in the MOJAVE sample and about one order of magnitude lower than those in flux-density-limited RFC subsample. Only three sources from our sample have a host-frame median core $T_\mathrm{b}$ higher than the equipartition value: J0741$+$3112, J1017$+$6116, and J1229$+$0203. Note that, as discussed above, J1229$+$0203 is an exceptional case for our sample. 

For the vast majority of our sample, the host-frame core $T_\mathrm{b}$ is lower than the equipartition value, and for about half of the sources of our sample with available redshifts, the difference is even an order of magnitude or more. For 13 sources in our full sample, no redhifts are provided in NED. Their observer's frame $T_\mathrm{b}$ ranges from $1.3\times10^8$~K to $6.0\times10^9$~K, so given the typical AGN redshifts, their host-frame $T_\mathrm{b}$ are likely also several times to two orders of magnitude lower than the equipartition brightness temperature. This is an indication for a very weak Doppler boosting and/or an extremely low intrinsic brightness temperature of the radio core for most of the sources in our final sample. The most massive estimates of AGN Doppler factors to date were published by the MOJAVE \citep{2021ApJ...923...67H} and OVRO \citep{2018ApJ...866..137L} teams. The former relies on the VLBI measurements of the core brightness temperature and the jet kinematics, and the latter -- on the variability timescale measurements. Only three sources of our sample have Doppler factors determined by any of these groups: J0741+3112, J0927+3902, and J1229+0203. They have significant Doppler factors, however, they may not be representative for the whole sample. At the same time, we do not detect the counter-jet in any of our sources, which implies a significant relativistic beaming. Therefore, the intrinsic brightness temperature is likely several times lower than the measured host-galaxy-frame values. Recently, \citet{2025A&A...695A.118B} also reported the radio core $T_\mathrm{b}$ lower than the equipartition value for a dozen AGNs. However, they observed at higher frequencies, where the core may become optically thin, and a number of their sources have two-sided jets, the presence of which indicates large viewing angles, in contrast with our sources.

\begin{figure}
    \centering
    \includegraphics[width=\columnwidth]{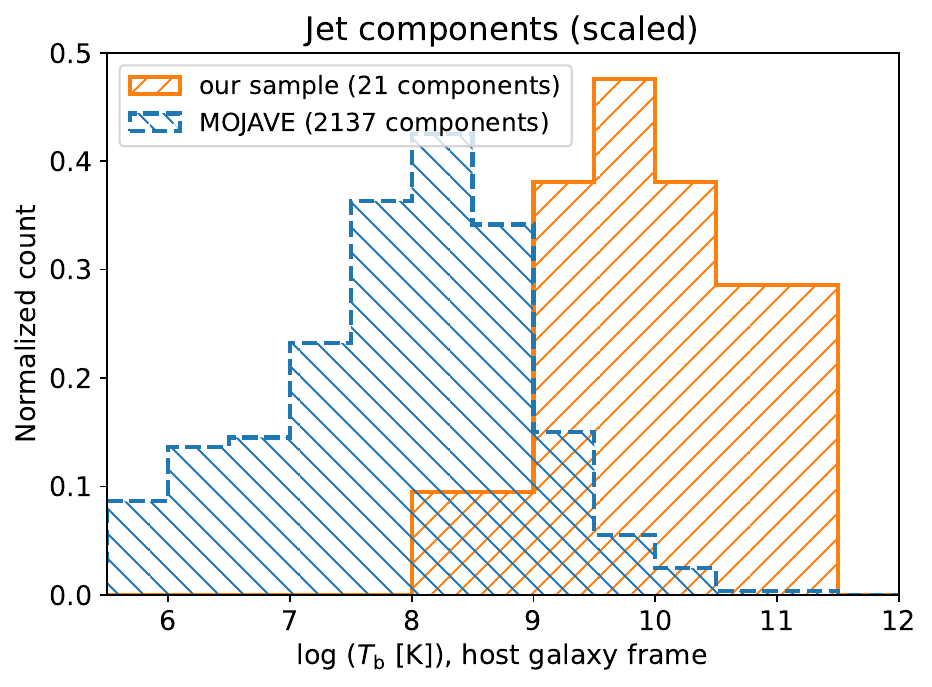}
    \caption{Empirical probability density histograms for the median of the logarithm of brightness temperatures of the dominant jet components in our sample (21 components), compared with the distribution for all the jet components of the sources in the MOJAVE 1.5~Jy Quarter Century sample \citep[][2137 components]{2021ApJ...923...30L}. The $T_\mathrm{b}$ values are given in the host galaxy reference frame (redshift correction applied, Doppler factor correction not applied) and are scaled to the same frequency and the same projected distance from the core for both samples -- see the details in \autoref{sec:res:tb}. After this scaling, the $T_\mathrm{b}$ values for the dominant jet components of our sources are about the highest observed in the MOJAVE project.}
    \label{fig:tb_jet}
\end{figure}

It is also interesting to check whether the dominant jet components, associated with the VLBI coordinates, have a typical or an enhanced brightness. For this reason, we made a comparison of our VLBI-associated components with all jet components of the AGNs from the MOJAVE 1.5~Jy Quarter Century sample. We calculated the host-galaxy-frame $T_\mathrm{b}$ of all the non-core components at all observing epochs for the sources with known redshifts using the components' flux densities and sizes published by \citet{2021ApJ...923...30L}. This results in 22514 brightness temperature measurements for 2137 components. Unlike the case of the cores, direct comparison of the brightness temperatures between the samples is incorrect, since the jet components' $T_\mathrm{b}$ strongly depends on the frequency (\autoref{fig:tb_spectra}) and on the distance from the core \citep{Kadlerphdthesis, 2012A&A...544A..34P, 2022A&A...660A...1B, 2025MNRAS.538.2008K}. For this reason, we scaled both our and MOJAVE values to the same frequency and the same projected distance from the core in the following way. Firstly, we used the $T_\mathrm{b}$ measurements for the sources of our sample made at 15~GHz, the MOJAVE frequency, or, if no 15~GHz data are available for a source, extrapolated the measurements at lower frequencies to 15~GHz. The linear extrapolation in the logarithmic scale was applied. Secondly, we scaled the resulting 15~GHz brightness temperatures of the jet components of both our and MOJAVE sources to the common reference linear projected distance $r$ from the core, using the relation $T_{\rm b}(r) \sim r^{-2.8}$ obtained by \citet{2025MNRAS.538.2008K}. The reference projected distance was set to 30~pc, which is the median projected distance between the core and the dominant jet component within our sample. The normalized distributions of the median $T_\mathrm{b}$ scaled in this way are shown in \autoref{fig:tb_jet}. For each jet component, one $T_\mathrm{b}$ value was used -- the median across all observing epochs. 
The distributions for the two samples are quite different: brightness temperatures of the dominant jet components in our sample belong to the right tail of the broad MOJAVE distribution (note the very different number of measurements in two samples). It should be taken into account that the jet components in our sample were selected in a very special way, being the brightest components of their jets. However, when we limit the MOJAVE distribution to only those jet components which have the maximum flux density after the core at the corresponding epochs, the distribution of the scaled $T_\mathrm{b}$ does not change significantly. Also, if we consider only the standing (``slow pattern speed'') MOJAVE components, again the peak of the distribution for our sample corresponds to the right tail of the MOJAVE distribution.

Therefore, the sources from our sample not only have remarkably faint radio cores, but also have extremely bright components in their radio jets. This is a result of the specific method of sample selection, as discussed in \autoref{sec:discus}.
Possible physical explanations for such an atypical structure will be discussed in \autoref{sec:discus:phys}.

\section{Discussion}
\label{sec:discus}

\subsection{Physical explanation for atypical brightness of cores and jet components}
\label{sec:discus:phys}

The astrophysical results from \autoref{sec:res} can be summarized as follows. The AGNs selected to our final sample have very dim radio jet cores with a brightness temperature on average an order of magnitude lower than the equipartition estimate. Several tens of parsecs away from the core, there is a jet component exceeding the core in brightness. There are indications that these bright jet components are stationary shocks.

The reason why our sample consists of sources having such an unusual structure is the specific method of selection. Namely, the sources whose parsec-scale radio structure is dominated by a bright and compact non-core detail were selected. Therefore, two peculiarities we see in our objects in combination -- the faint core and the bright jet component -- may be physically connected or not connected. There are several possible scenarios that may explain the properties of these sources.

One of possible clues to understanding the physical conditions in our sources is their similarity with one well-studied example, the quasar PKS~0858$-$279, also known as J0900$-$2808 \citep{2022MNRAS.510.1480K, 2024MNRAS.528.1697K}. The multi-frequency, multi-epoch polarimetric VLBI observations of this gigahertz-peaked spectrum source showed that its dominating feature is a bright component of the jet at a projected distance of 20~pc from the core. A jet bending occurs at the position of this feature. Using the results of the model fitting published by \citet{2024MNRAS.528.1697K}, we calculated the brightness temperatures of the core and the dominant jet component. The host-galaxy-frame core $T_\mathrm{b}$ varies from $8\times10^{8}$~K to $4\times10^{10}$~K with a median of $1.2\times10^{10}$~K. The dominant jet component has a minimum $T_\mathrm{b}$ of $1\times10^{10}$~K, a maximum of $6\times10^{10}$~K, and a median of $3.5\times10^{10}$~K. These values are quite similar to those for typical sources in our sample. A number of our sources also share other properties of PKS~0858$-$279, such as a bended jet, a peaked spectral shape, and the QSO (quasi-stellar object, or quasar) optical class. In particular, a visual analysis of the maps shows that in 13 sources from our full sample, the jet bends in a region of the VLBI-associated component or near it. 
In 11 more sources, the jet on the available maps is not detected at distances further from the core than the VLBI-associated component. Therefore, we cannot exclude the jet bending in these 11 sources too; future, more sensitive VLBI observations may detect it.
Furthermore, the inspection of the integral broadband radio spectra from the CATS database\footnote{\url{https://www.sao.ru/cats/}} \citep{2005BSAO...58..118V} revealed that the peaked-spectrum sources account for about a half of our sample. This is a very high fraction: for comparison, the fraction of the gigahertz-peaked spectrum sources in complete flux-density-limited samples is about 1-5\% \citep[e.~g.,][]{2009AN....330..199S, 2011ARep...55..187M, 2013AstBu..68..262M, 2019AstBu..74..348S}. 
Finally, 17 of 24 sources of our sample with optical identifications in the NED database are quasars (QSO), like PKS~0858$-$279.
It is worth mentioning that PKS~0858$-$279 was selected by our algorithm as a candidate at the first step of our search (\autoref{sec:select:init}). It was not selected for our final sample just because the lack of high-frequency data for this object in the Astrogeo database did not allow us to measure the spectral indices of its components. \citet{2022MNRAS.510.1480K, 2024MNRAS.528.1697K} provided evidence that the cause of the stationary jet brightening in PKS~0858$-$279 is the interaction of the jet with a dense interstellar medium, which leads to a shock wave formation. As was mentioned above, the dominating jet components in at least a part of our sources are probably also the standing shock fronts. Therefore, the same mechanism can work in them. However, it does not explain the reduced brightness of the core.

We consider several explanations of the extremely low core brightness temperatures in the AGNs in our sample. The first one is the deviation from the energy equipartition to the magnetic field energy dominance in the core region. This may be connected with ineffective particles heating. In this case, the transformation of the magnetic energy into the particles energy, i.~e., particle acceleration, can occur further along the jet, which causes the formation of the bright jet component. This allows to explain the reduced core brightness and enhanced jet component brightness by the same scenario. Alternatively, the core may be located in a region where the jet itself is still weakly accelerated and magnetically dominated. This is more often the case at higher frequencies \citep{2025A&A...695A.118B}, but if the synchrotron self-absorption (SSA) is weaker in our sources than usual for some reasons, the radio core at the studied frequencies (mostly 2--8~GHz) may in principle lie in the magnetically dominated region.

Secondly, the faintness of the cores could be explained by the differential Doppler boosting \citep{1994cers.conf..245K}. Traditionally used estimate of the boosting factor $\delta^{n - \alpha}$ with $n = 2$ or $3$, where $\delta$ is the Doppler factor, is applicable for a power-law spectrum. However, a homogeneous synchrotron source has the spectrum with a maximum that divides the optically thick part with $\alpha = 2.5$ at lower frequencies and the optically thin part at higher frequencies. In the optically thick part, the shift of the spectrum approximately compensates for the Doppler boosting or even overcomes it for the continuous flow with $n = 2$ \citep{1994cers.conf..245K}. In principle, the source does not have to be homogeneous to display such an effect. For example, it will also be present if the emitting particles are heated only in some narrow range of distances from the jet origin and cooled very efficiently. However, such a scenario implies the absence of a significant frequency-dependent core shift. 

Thirdly, the strong absorption, either SSA or free-free, may decrease the core brightness. For example, the free-free absorption may be quite strong in the presence of a dense surrounding medium, like in PKS~0858$-$279. Several sources from our sample demonstrate significantly inverted core spectra, \mbox{$\alpha_\mathrm{core}>+1$} (see \autoref{tab:sample} and the full version of \autoref{tab:pairs}), which may indicate substantial absorption. However, the spectral index errors are quite large for most of them. Future observations are needed to measure the core spectra more precisely and study the role of absorption in decreasing their core brightness.

The information we have at the moment is not enough to choose the most probable physical explanation for the reduced core brightness and the enhanced brightness of the jet component in our sources. In 2024-2025, we ran a dedicated VLBA campaign to observe several sources from this sample with better sensitivity and frequency coverage in the full-polarization mode. We expect that these new data will provide information on the magnetic field strength and configuration, as well as on the absorption properties of these objects, which will make it possible to understand the physical origin of their exotic brightness distribution.

\subsection{Degree of sample completeness}
\label{sec:discus:completeness}

\begin{figure*}
    \centering
    \includegraphics[width=0.49\linewidth]{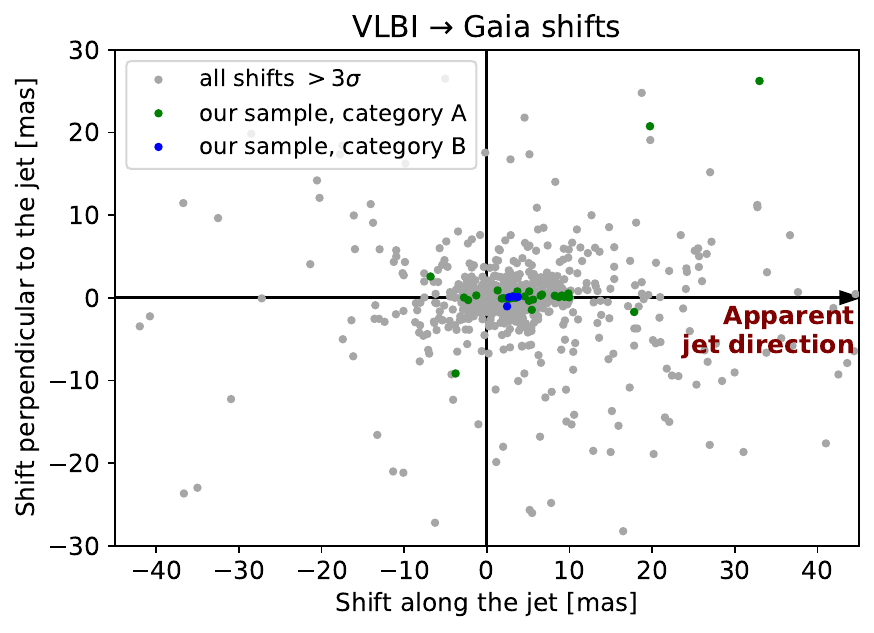}
    \includegraphics[width=0.49\linewidth]{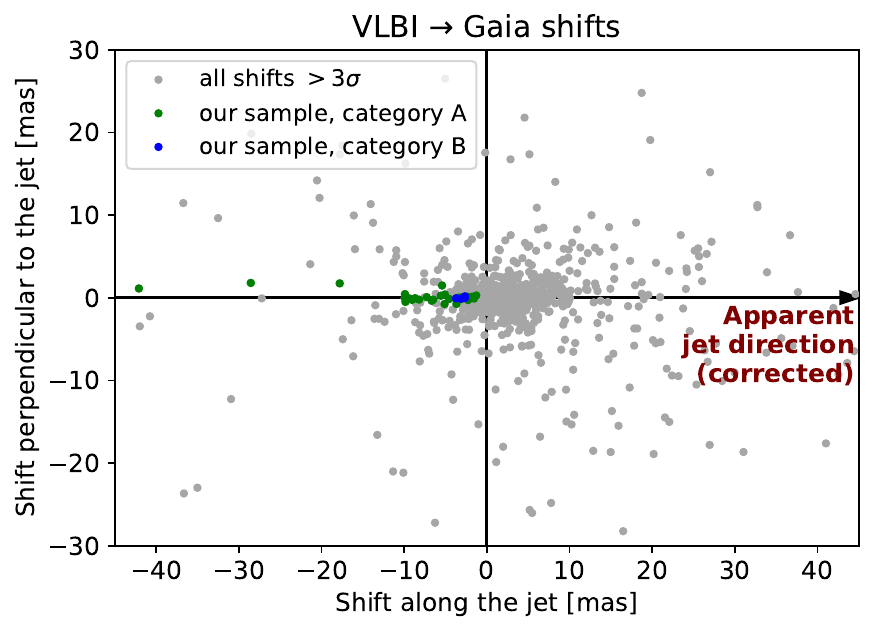}
    \caption{Diagram showing the statistically significant ($>3\sigma$) \vg shifts with respect to the jet direction from \citet{2022ApJS..260....4P} ({\it left}) and the updated jet direction after the corrections (see \autoref{sec:res:jetdir} and \autoref{tab:sample}) applied ({\it right}). The jet direction corresponds to the positive direction of the horizontal axis. The shifts for the sources from our clean sample (category A) are marked by green dots, for the sources from the full sample not belonging to the clean sample (category B)~-- by blue dots. Note that, after the correction, the number of known large \vg shifts opposite to the jet direction grows significantly.}
    \label{fig:diagram}
\end{figure*}

To show the sample we selected in the context of all sources with statistically significant \vg shifts, we plot all the sources with the \vg shift $>3\sigma_\mathrm{vg}$ in \autoref{fig:diagram}, highlighting the sources of our sample by colour. The coordinates of this plot are the projections of the \vg shift to the line of the jet direction and to the line perpendicular to it. To recall, the \vg shift is the vector with the origin at the VLBI absolute position and the head at the \g absolute position of a source throughout our work. The jet directions for all sources in the left panel of \autoref{fig:diagram} and for most sources in its right panel were taken from \citet{2022ApJS..260....4P}. As discussed in \autoref{sec:res:jetdir}, the jet direction automatically determined by \citet{2022ApJS..260....4P} must be reversed or significantly corrected for most of the sources in our sample because the core was previously misidentified for them. For the right panel of \autoref{fig:diagram}, these corrections were applied. One can see that our sample is a minority of all sources with statistically significant \vg shifts. However, after the correction of the jet direction, our sources now occupy the previously sparsely populated region in \autoref{fig:diagram} of large shifts opposite to the jet direction. Three of them, J1630$-$1157, J1821+3945, and J0823$-$0939, have remarkably large \vg shifts (see also \autoref{tab:sample}).

The share of our sample among all known AGNs with statistically significant \vg shifts can easily be calculated. We found \vg shifts with statistical significance \mbox{$>3\sigma$} and length~\mbox{$>1$}~mas for 2167 sources.
Of them, we selected our clean sample of 31 sources and our full sample of 35 sources. This gives the fractions of 1.4\% and 1.6\%, correspondingly. Among sources with \vg shifts larger than 1 mas and with a significance higher than 5$\sigma$, the sources of our clean sample account for 2.2\%, and of our full sample~-- for 2.5\%. 

However, we stress that our sample cannot be considered complete for several reasons. Firstly, we apply rather strict filtering by errors at the step of the initial search for candidates (\autoref{sec:select:init}). This is important for an automatic search in a large database containing the data of very different quality, but also leads to the unavoidable loss of some candidates. Secondly, we consider only the \vg{} shifts larger than 1~mas for the reasons described in \autoref{sec:select:init}. Thirdly, the available public VLBI images allowed us to conduct the full analysis, including the measurement of the spectral indices, for only about a one-third of the initial candidates. And finally, the situation when a bright non-core feature dominates the parsec-scale structure does not necessarily lead to the alignment between the \vg{} shift and the vector between two brightest components for the reasons we discuss below.

We inspected the sources for which it was reported by different authors that their VLBI coordinates are associated with a non-core component at some bands and which, however, were not selected to our sample. We found that these cases fall into one of the following groups.

\begin{itemize}
    \item The source was selected as a candidate, but the available open-access VLBI images did not allow us to measure the spectral indices of the components. The examples of this case are the above-mentioned J0900$-$2808 \citep[PKS~0858$-$279;][]{2022MNRAS.510.1480K, 2024MNRAS.528.1697K} and also J1217+5835 \citep{2024AJ....168...76P}.
    \item The shift between the position provided by the RFC fused solution, used as the VLBI position throughout our work, and the \g{} position is less than 1~mas, which excludes from our consideration the sources like J0745$-$0044 \citep[0743$-$006;][]{2021A&A...647A.189X}, which physically are similar to the sources of our sample.
    \item The radio core associated with the \g{} coordinates is even fainter than in typical sources from our sample, and for this reason the two-component brightness distribution models, used in our search for candidates, do not include the core component. This is the case of sources like J0432+4138 \citep[3C 119;][]{1991A&A...245..449N, 2011AJ....142...35P, 2021A&A...647A.189X} or J1330+2509 \citep[1328+254 or 3C 287;][]{2022MNRAS.512..874T, 2025arXiv250106513F}. A more extreme case is if the core is so faint that it has not yet been detected by the VLBI.
    \item Since different structure components dominate at different bands, the fused VLBI solution provides the position in the middle of the components, like in J0725$-$0054 \citep[0723$-$008;][]{2025AJ....169..173X} or J1159$-$2148 \citep[1157$-$215;][]{2022A&A...663A..83X}.
    \item The source has a morphology of the compact symmetric object \citep[CSO; e.g.,][]{1994ApJ...432L..87W, 2011A&A...535A..24S, 2024ApJ...961..240K}. In this case, two dominant VLBI features are the hot spots or mini-lobes at two sides of the jet origin, and the VLBI coordinates are associated with one of them. The \g{} coordinates, presumably associated with the central engine, lie between two dominant radio features. The \vg{} shifts are also very informative for searching such objects, but a different approach has to be used, e.~g., like the one implemented in the recent work by \citet{2025arXiv250307288A}. Other individual CSOs were also found using the \vg{} shifts analysis by \citet{2020MNRAS.496.1811K} (J1110+4817) and  \citet{2025arXiv250106513F} (J1203+0414).
    \item Finally, one of the basic assumptions of our search algorithm -- that the \g{} coordinates match the jet base -- breaks in sources that demonstrate strong extended optical jets \citep{2017A&A...598L...1K, 2019ApJ...871..143P}. In this case, the position of the \g{} centroid is not linked directly to the radio structure features. 
\end{itemize}

Taking into account all these cases, it is obvious that the fraction of sources similar to those in our sample among all AGNs with significant \vg shifts should be higher than 1-2\% calculated above. More thorough studies are needed to create a more complete sample of them.

Note also that though the \vg shifts analysis is an efficient tool for finding AGNs with faint radio cores and bright jets, these shifts correspond to the vector between the dominant jet feature and the core not in all such sources. That is, the study of the radio jets with dim cores is a broader problem than just the interpretation of the \vg shifts. In fact, there are at least two morphological classes of compact radio-loud AGNs in which the core does not dominate the VLBI structure: CSOs and the objects like those from our sample~-- core-jet sources with jet components brighter than the core. The sources of both classes often have a steep or peaked spectrum in the GHz range, in contrast to the core-dominated sources having a flat spectrum. The relative prevalence of these two classes is a subject of future studies. However, it is already clear that together they account for a large fraction of compact AGNs. Namely, about one-third of the AGNs in the current version of the RFC catalogue have VLBI spectral indices $<-0.5$ in the 5-8~GHz range. Our VLBA survey of a statistically complete, flux-density-limited sample of about 500 AGNs near the North Celestial Pole \citep{2021AJ....161...88P} demonstrated that about a half of compact sources have steep or peaked spectra in the 2--8~GHz range. All these AGNs should have the non-core-dominated parsec-scale structure.


\subsection{Outlook}

As a result of our systematic search, we found 35 sources for which the astrometric solutions in different bands pinpoint at different structure details separated by several to tens of mas. Such sources must be treated with caution in any astrometric analysis. Firstly, these objects should not be used for the comparison of the VLBI and \g celestial reference frames, since VLBI and \g measure the coordinates of their different regions. Secondly, the different spectra of these two components can lead to a situation when the steep-spectrum jet component dominates at lower VLBI frequencies, while the flat-spectrum radio core -- at higher frequencies, as we demonstrated for the source J0741+3112 from our sample, similarly to the examples found previously by \citet{2011AJ....142...35P, 2022A&A...663A..83X, 2024AJ....168...76P, 2025AJ....169..173X}. Therefore, it is essential to account for sources like those in our sample when conducting multi-frequency VLBI astrometry, particularly across a broad frequency range \citep{2024AJ....168...76P, 2024AJ....168..219X}. Thirdly, \citet{2022MNRAS.512..874T} and \citet{2024AstL...50..657O} reported abrupt changes in the positions of a number of sources, most of which have two compact components in their VLBI images. Variability of the brightness of the components can change their dominance, and therefore the positions at different epochs will correspond to different components. Although this phenomenon was not observed in the sources from our sample, we cannot rule out its occurrence in the future. 

For all these reasons, AGNs like those in our sample should be flagged in astrometric catalogues and their peculiarity should be taken into account. For applications that require stability, for instance, phase referencing or geodesy, the use of sources with two or more compact components is undesirable. On the other hand, continuous astrometric observations of these sources have the potential to yield interesting scientific results. Moreover, as was pointed out in \autoref{sec:discus:phys}, deeper astrophysical study of such sources, having dim cores and bright jet components, is necessary to find the most plausible physical explanation for their remarkable properties. The new observations we made for several sources from our sample are currently in the data reduction stage.

All the considerations above also show that further search for AGNs with dominating non-core features in their compact radio jets is an important problem. Although the analysis of the \vg shifts has proven to be an effective tool for finding such sources, this type of radio structure does not necessarily cause a significant \vg offset. A more general approach is to analyse the spectral indices of the major compact components for all the sources used in the VLBI astrometry and geodesy. Most of such observations are dual-frequency or multi-frequency, so the spectral index analysis in a fashion similar to ours is possible for them.


\section{Summary}
\label{sec:sum}

We performed a systematic search for the AGNs whose VLBI and \g coordinates correspond to their different parsec-scale radio structure components. We used the publicly available Radio Fundamental Catalogue, \g DR3, and the Astrogeo VLBI image database. We identified the so-called radio core on the VLBI maps by its flat spectrum. 
Our main results are summarized as follows.

\begin{enumerate}
    \item We constructed a sample of AGNs whose VLBI coordinates do not correspond to the apparent jet origin, but to a feature in the radio jet parsecs away from the radio core. At the same time, their optical \g coordinates coincide with the jet origin and are close to the radio core, as expected. The sample consists of 35 sources, for 31 of which the identification of the core is the most reliable.
    \item These effects on VLBI positions are caused by radio cores persistently being unusually weak and other jet components persistently being unusually bright at the same time. 
    \item The cores of practically all of the sources of this sample have brightness temperatures in the host galaxy reference frame lower than the equipartition value of $\approx5\times10^{10}$~K. This challenges the models of the energy balance and the particle acceleration in their jets or requires some special absorption regime.
    \item The non-core jet components dominating the VLBI structure have spectral indices in the range $-2\lesssim\alpha\lesssim0$. For most of our sources, these components probably are quasi-stationary shocks or reconnection layers \citep{1985ApJ...298..114M, 2013MNRAS.436.3341Z}. The indications of this are the $\nu^{-0.5}$ dependence of size on frequency and the absence of the excess absolute VLBI proper motion. These shocks may arise as a result of the interaction with the surrounding medium or for some other reasons.
    \item The unusual morphology of the sources in our sample necessitates special treatment in the following aspects. They should not be used as reference sources for the celestial reference frame. The multi-band VLBI astrometry should account for the fact that their absolute VLBI coordinates can differ significantly at different frequencies, with the coordinates at higher frequencies matching the core as well as the \g coordinates, while the coordinates at lower frequencies matching the brightest jet detail. The jet directions previously determined by \citet{2022ApJS..260....4P} must be reversed for most of them. 
\end{enumerate}

Our study has demonstrated that AGNs with faint radio cores and bright jet features are numerous, complicate the multi-band astrometric analysis, but at the same time provide valuable material for tests of the physical models of the relativistic jet emission. 
Further study of such sources and the construction of their more complete sample are important tasks.


\section*{Acknowledgements}

We thank Alexander Pushkarev, Minghui Xu, and Elena Nokhrina for fruitful discussions on this project. 
We thank the anonymous referee and Eduardo Ros for comments on the manuscript that
helped us to improve this paper.
This work was supported by the Russian Science Foundation project 20-72-10078.
A.~V.~Plavin is a postdoctoral fellow at
the Black Hole Initiative, which is funded by grants from the John Templeton Foundation (grants 60477, 61479, 62286) and the Gordon and Betty Moore Foundation (grant GBMF-8273). 
YYK is supported by the MuSES project, which has received funding from the European Union (ERC grant agreement No 101142396).
%
The views and opinions expressed in this work are those of the authors and do not necessarily reflect the views of these Foundations.

We acknowledge the use of the Radio Fundamental Catalogue \citep{RFC}.
We used in our work the Astrogeo VLBI FITS image database, maintained by Leonid Petrov.
This research made use of the data from the MOJAVE database maintained by the MOJAVE team \citep{2018ApJS..234...12L}.
This work has made use of data from the European Space Agency (ESA) mission {\it Gaia} (\url{https://www.cosmos.esa.int/gaia}), processed by the {\it Gaia} Data Processing and Analysis Consortium (DPAC, \url{https://www.cosmos.esa.int/web/gaia/dpac/consortium}). Funding for the DPAC has been provided by national institutions, in particular the institutions participating in the {\it Gaia} Multilateral Agreement.
This research has made use of the NASA/IPAC Extragalactic Database, which is funded by the National Aeronautics and Space Administration and operated by the California Institute of Technology.
This research has made use of NASA's Astrophysics Data System Bibliographic Services.
This research has made use of the CATS database, operated at SAO RAS, Russia \citep{2005BSAO...58..118V}.
This work made use of Astropy\footnote{\url{http://www.astropy.org}}, a community-developed core Python package and an ecosystem of tools and resources for astronomy \citep{2022ApJ...935..167A}.

\section*{Data Availability}

The data underlying this article are publicly available 
in the Radio Fundamental Catalogue 
\href{https://doi.org/10.25966/dhrk-zh08}{DOI:10.25966/dhrk-zh08},
in the Astrogeo VLBI FITS image database 
\href{https://doi.org/10.25966/kyy8-yp57}{DOI:10.25966/kyy8-yp57},
and in the \g Data Release 3 at \url{https://gea.esac.esa.int/archive/}.


\bibliographystyle{mnras}
\bibliography{VLBI-Gaia-2comp}

\begin{thebibliography}{}
\makeatletter
\relax
\def\mn@urlcharsother{\let\do\@makeother \do\$\do\&\do\#\do\^\do\_\do\%\do\~}
\def\mn@doi{\begingroup\mn@urlcharsother \@ifnextchar [ {\mn@doi@} {\mn@doi@[]}}
\def\mn@doi@[#1]#2{\def\@tempa{#1}\ifx\@tempa\@empty \href {http://dx.doi.org/#2} {doi:#2}\else \href {http://dx.doi.org/#2} {#1}\fi \endgroup}
\def\mn@eprint#1#2{\mn@eprint@#1:#2::\@nil}
\def\mn@eprint@arXiv#1{\href {http://arxiv.org/abs/#1} {{\tt arXiv:#1}}}
\def\mn@eprint@dblp#1{\href {http://dblp.uni-trier.de/rec/bibtex/#1.xml} {dblp:#1}}
\def\mn@eprint@#1:#2:#3:#4\@nil{\def\@tempa {#1}\def\@tempb {#2}\def\@tempc {#3}\ifx \@tempc \@empty \let \@tempc \@tempb \let \@tempb \@tempa \fi \ifx \@tempb \@empty \def\@tempb {arXiv}\fi \@ifundefined {mn@eprint@\@tempb}{\@tempb:\@tempc}{\expandafter \expandafter \csname mn@eprint@\@tempb\endcsname \expandafter{\@tempc}}}

\bibitem[\protect\citeauthoryear{{Albareti} et~al.,}{{Albareti} et~al.}{2017}]{2017ApJS..233...25A}
{Albareti} F.~D.,  et~al., 2017, \mn@doi [\apjs] {10.3847/1538-4365/aa8992}, \href {https://ui.adsabs.harvard.edu/abs/2017ApJS..233...25A} {233, 25}

\bibitem[\protect\citeauthoryear{{An}, {Zhang}, {Frey}, {Baan}  \& {Wang}}{{An} et~al.}{2025}]{2025arXiv250307288A}
{An} T.,  {Zhang} Y.,  {Frey} S.,  {Baan} W.~A.,   {Wang} A.,  2025, \mn@doi [arXiv e-prints] {10.48550/arXiv.2503.07288}, \href {https://ui.adsabs.harvard.edu/abs/2025arXiv250307288A} {p. arXiv:2503.07288}

\bibitem[\protect\citeauthoryear{{Astropy Collaboration} et~al.,}{{Astropy Collaboration} et~al.}{2022}]{2022ApJ...935..167A}
{Astropy Collaboration} et~al., 2022, \mn@doi [\apj] {10.3847/1538-4357/ac7c74}, \href {https://ui.adsabs.harvard.edu/abs/2022ApJ...935..167A} {935, 167}

\bibitem[\protect\citeauthoryear{{Beasley}, {Gordon}, {Peck}, {Petrov}, {MacMillan}, {Fomalont}  \& {Ma}}{{Beasley} et~al.}{2002}]{r:vcs1}
{Beasley} A.~J.,  {Gordon} D.,  {Peck} A.~B.,  {Petrov} L.,  {MacMillan} D.~S.,  {Fomalont} E.~B.,   {Ma} C.,  2002, \mn@doi [\apjs] {10.1086/339806}, \href {http://adsabs.harvard.edu/cgi-bin/nph-bib_query?bibcode=2002ApJS..141...13B&db_key=AST} {141, 13}

\bibitem[\protect\citeauthoryear{{Blandford} \& {K{\"o}nigl}}{{Blandford} \& {K{\"o}nigl}}{1979}]{1979ApJ...232...34B}
{Blandford} R.~D.,  {K{\"o}nigl} A.,  1979, \mn@doi [\apj] {10.1086/157262}, \href {https://ui.adsabs.harvard.edu/abs/1979ApJ...232...34B} {232, 34}

\bibitem[\protect\citeauthoryear{{Boccardi} et~al.,}{{Boccardi} et~al.}{2025}]{2025A&A...695A.118B}
{Boccardi} B.,  et~al., 2025, \mn@doi [\aap] {10.1051/0004-6361/202453138}, \href {https://ui.adsabs.harvard.edu/abs/2025A&A...695A.118B} {695, A118}

\bibitem[\protect\citeauthoryear{{Burd}, {Kadler}, {Mannheim}, {Baczko}, {Ringholz}  \& {Ros}}{{Burd} et~al.}{2022}]{2022A&A...660A...1B}
{Burd} P.~R.,  {Kadler} M.,  {Mannheim} K.,  {Baczko} A.~K.,  {Ringholz} J.,   {Ros} E.,  2022, \mn@doi [\aap] {10.1051/0004-6361/202142363}, \href {https://ui.adsabs.harvard.edu/abs/2022A&A...660A...1B} {660, A1}

\bibitem[\protect\citeauthoryear{{Burke-Spolaor}}{{Burke-Spolaor}}{2011}]{2011MNRAS.410.2113B}
{Burke-Spolaor} S.,  2011, \mn@doi [\mnras] {10.1111/j.1365-2966.2010.17586.x}, \href {https://ui.adsabs.harvard.edu/abs/2011MNRAS.410.2113B} {410, 2113}

\bibitem[\protect\citeauthoryear{{Cao}, {Zheng}, {Biesiada}, {Qi}, {Chen}  \& {Zhu}}{{Cao} et~al.}{2017}]{2017A&A...606A..15C}
{Cao} S.,  {Zheng} X.,  {Biesiada} M.,  {Qi} J.,  {Chen} Y.,   {Zhu} Z.-H.,  2017, \mn@doi [\aap] {10.1051/0004-6361/201730551}, \href {https://ui.adsabs.harvard.edu/abs/2017A&A...606A..15C} {606, A15}

\bibitem[\protect\citeauthoryear{{Charlot}}{{Charlot}}{2002}]{2002ivsg.conf..233C}
{Charlot} P.,  2002, in {Vandenberg} N.~R.,  {Baver} K.~D.,  eds, International VLBI Service for Geodesy and Astrometry: General Meeting Proceedings. p.~233

\bibitem[\protect\citeauthoryear{{Charlot} et~al.,}{{Charlot} et~al.}{2020}]{2020A&A...644A.159C}
{Charlot} P.,  et~al., 2020, \mn@doi [\aap] {10.1051/0004-6361/202038368}, \href {https://ui.adsabs.harvard.edu/abs/2020A&A...644A.159C} {644, A159}

\bibitem[\protect\citeauthoryear{{Chen}, {Liu}, {Lazio}, {Breiding}, {Burke-Spolaor}, {Hwang}, {Shen}  \& {Zakamska}}{{Chen} et~al.}{2023}]{2023ApJ...958...29C}
{Chen} Y.-C.,  {Liu} X.,  {Lazio} J.,  {Breiding} P.,  {Burke-Spolaor} S.,  {Hwang} H.-C.,  {Shen} Y.,   {Zakamska} N.~L.,  2023, \mn@doi [\apj] {10.3847/1538-4357/ad00b3}, \href {https://ui.adsabs.harvard.edu/abs/2023ApJ...958...29C} {958, 29}

\bibitem[\protect\citeauthoryear{{Cheng} et~al.,}{{Cheng} et~al.}{2020}]{2020ApJS..247...57C}
{Cheng} X.~P.,  et~al., 2020, \mn@doi [\apjs] {10.3847/1538-4365/ab791f}, \href {https://ui.adsabs.harvard.edu/abs/2020ApJS..247...57C} {247, 57}

\bibitem[\protect\citeauthoryear{{Croom}, {Smith}, {Boyle}, {Shanks}, {Miller}, {Outram}  \& {Loaring}}{{Croom} et~al.}{2004}]{2004MNRAS.349.1397C}
{Croom} S.~M.,  {Smith} R.~J.,  {Boyle} B.~J.,  {Shanks} T.,  {Miller} L.,  {Outram} P.~J.,   {Loaring} N.~S.,  2004, \mn@doi [\mnras] {10.1111/j.1365-2966.2004.07619.x}, \href {https://ui.adsabs.harvard.edu/abs/2004MNRAS.349.1397C} {349, 1397}

\bibitem[\protect\citeauthoryear{{DESI Collaboration} et~al.,}{{DESI Collaboration} et~al.}{2024}]{2024AJ....168...58D}
{DESI Collaboration} et~al., 2024, \mn@doi [\aj] {10.3847/1538-3881/ad3217}, \href {https://ui.adsabs.harvard.edu/abs/2024AJ....168...58D} {168, 58}

\bibitem[\protect\citeauthoryear{{De Rosa} et~al.,}{{De Rosa} et~al.}{2019}]{2019NewAR..8601525D}
{De Rosa} A.,  et~al., 2019, \mn@doi [\nar] {10.1016/j.newar.2020.101525}, \href {https://ui.adsabs.harvard.edu/abs/2019NewAR..8601525D} {86, 101525}

\bibitem[\protect\citeauthoryear{{Fey} et~al.,}{{Fey} et~al.}{2015}]{r:icrf2}
{Fey} A.~L.,  et~al., 2015, \mn@doi [\aj] {10.1088/0004-6256/150/2/58}, \href {https://ui.adsabs.harvard.edu/abs/2015AJ....150...58F} {150, 58}

\bibitem[\protect\citeauthoryear{{Frey}, {Titov}, {Melnikov}  \& {Lambert}}{{Frey} et~al.}{2025}]{2025arXiv250106513F}
{Frey} S.,  {Titov} O.,  {Melnikov} A.,   {Lambert} S.,  2025, \mn@doi [arXiv e-prints] {10.48550/arXiv.2501.06513}, \href {https://ui.adsabs.harvard.edu/abs/2025arXiv250106513F} {p. arXiv:2501.06513}

\bibitem[\protect\citeauthoryear{{Gaia Collaboration} et~al.,}{{Gaia Collaboration} et~al.}{2016}]{2016A&A...595A...1G}
{Gaia Collaboration} et~al., 2016, \mn@doi [\aap] {10.1051/0004-6361/201629272}, \href {https://ui.adsabs.harvard.edu/abs/2016A&A...595A...1G} {595, A1}

\bibitem[\protect\citeauthoryear{{Gaia Collaboration} et~al.,}{{Gaia Collaboration} et~al.}{2018}]{2018A&A...616A..14G}
{Gaia Collaboration} et~al., 2018, \mn@doi [\aap] {10.1051/0004-6361/201832916}, \href {https://ui.adsabs.harvard.edu/abs/2018A&A...616A..14G} {616, A14}

\bibitem[\protect\citeauthoryear{{Gaia Collaboration} et~al.,}{{Gaia Collaboration} et~al.}{2022}]{2022A&A...667A.148G}
{Gaia Collaboration} et~al., 2022, \mn@doi [\aap] {10.1051/0004-6361/202243483}, \href {https://ui.adsabs.harvard.edu/abs/2022A&A...667A.148G} {667, A148}

\bibitem[\protect\citeauthoryear{{Gaia Collaboration} et~al.,}{{Gaia Collaboration} et~al.}{2023}]{2023A&A...674A...1G}
{Gaia Collaboration} et~al., 2023, \mn@doi [\aap] {10.1051/0004-6361/202243940}, \href {https://ui.adsabs.harvard.edu/abs/2023A&A...674A...1G} {674, A1}

\bibitem[\protect\citeauthoryear{{Gupta} et~al.,}{{Gupta} et~al.}{2022}]{2022ApJ...929..108G}
{Gupta} N.,  et~al., 2022, \mn@doi [\apj] {10.3847/1538-4357/ac4220}, \href {https://ui.adsabs.harvard.edu/abs/2022ApJ...929..108G} {929, 108}

\bibitem[\protect\citeauthoryear{{Henstock}, {Browne}, {Wilkinson}  \& {McMahon}}{{Henstock} et~al.}{1997}]{1997MNRAS.290..380H}
{Henstock} D.~R.,  {Browne} I.~W.~A.,  {Wilkinson} P.~N.,   {McMahon} R.~G.,  1997, \mn@doi [\mnras] {10.1093/mnras/290.2.380}, \href {https://ui.adsabs.harvard.edu/abs/1997MNRAS.290..380H} {290, 380}

\bibitem[\protect\citeauthoryear{{Homan} et~al.,}{{Homan} et~al.}{2021}]{2021ApJ...923...67H}
{Homan} D.~C.,  et~al., 2021, \mn@doi [\apj] {10.3847/1538-4357/ac27af}, \href {https://ui.adsabs.harvard.edu/abs/2021ApJ...923...67H} {923, 67}

\bibitem[\protect\citeauthoryear{{Hovatta} et~al.,}{{Hovatta} et~al.}{2014}]{2014AJ....147..143H}
{Hovatta} T.,  et~al., 2014, \mn@doi [\aj] {10.1088/0004-6256/147/6/143}, \href {https://ui.adsabs.harvard.edu/abs/2014AJ....147..143H} {147, 143}

\bibitem[\protect\citeauthoryear{Kadler}{Kadler}{2005}]{Kadlerphdthesis}
Kadler M.,  2005, PhD thesis, Rheinische Friedrich-Wilhelms-Universit{\"a}t, Bonn, Germany

\bibitem[\protect\citeauthoryear{{Kellermann}}{{Kellermann}}{1994}]{1994cers.conf..245K}
{Kellermann} K.~I.,  1994, in {Zensus} J.~A.,  {Kellermann} K.~I.,  eds, Compact Extragalactic Radio Sources. National Radio Astronomy Observatory (NRAO), Green Bank, WV, p.~245

\bibitem[\protect\citeauthoryear{{Kiehlmann} et~al.,}{{Kiehlmann} et~al.}{2024}]{2024ApJ...961..240K}
{Kiehlmann} S.,  et~al., 2024, \mn@doi [\apj] {10.3847/1538-4357/ad0c56}, \href {https://ui.adsabs.harvard.edu/abs/2024ApJ...961..240K} {961, 240}

\bibitem[\protect\citeauthoryear{{Koryukova}, {Pushkarev}, {Plavin}  \& {Kovalev}}{{Koryukova} et~al.}{2022}]{2022MNRAS.515.1736K}
{Koryukova} T.~A.,  {Pushkarev} A.~B.,  {Plavin} A.~V.,   {Kovalev} Y.~Y.,  2022, \mn@doi [\mnras] {10.1093/mnras/stac1898}, \href {https://ui.adsabs.harvard.edu/abs/2022MNRAS.515.1736K} {515, 1736}

\bibitem[\protect\citeauthoryear{{Kosogorov}, {Kovalev}, {Perucho}  \& {Kovalev}}{{Kosogorov} et~al.}{2022}]{2022MNRAS.510.1480K}
{Kosogorov} N.~A.,  {Kovalev} Y.~Y.,  {Perucho} M.,   {Kovalev} Y.~A.,  2022, \mn@doi [\mnras] {10.1093/mnras/stab3579}, \href {https://ui.adsabs.harvard.edu/abs/2022MNRAS.510.1480K} {510, 1480}

\bibitem[\protect\citeauthoryear{{Kosogorov}, {Kovalev}, {Perucho}  \& {Kovalev}}{{Kosogorov} et~al.}{2024}]{2024MNRAS.528.1697K}
{Kosogorov} N.~A.,  {Kovalev} Y.~Y.,  {Perucho} M.,   {Kovalev} Y.~A.,  2024, \mn@doi [\mnras] {10.1093/mnras/stae084}, \href {https://ui.adsabs.harvard.edu/abs/2024MNRAS.528.1697K} {528, 1697}

\bibitem[\protect\citeauthoryear{{Kovalev} et~al.,}{{Kovalev} et~al.}{2005}]{2005AJ....130.2473K}
{Kovalev} Y.~Y.,  et~al., 2005, \mn@doi [\aj] {10.1086/497430}, \href {https://ui.adsabs.harvard.edu/abs/2005AJ....130.2473K} {130, 2473}

\bibitem[\protect\citeauthoryear{{Kovalev}, {Lobanov}, {Pushkarev}  \& {Zensus}}{{Kovalev} et~al.}{2008}]{2008AA...483..759K}
{Kovalev} Y.~Y.,  {Lobanov} A.~P.,  {Pushkarev} A.~B.,   {Zensus} J.~A.,  2008, \mn@doi [\aap] {10.1051/0004-6361:20078679}, \href {https://ui.adsabs.harvard.edu/abs/2008A&A...483..759K} {483, 759}

\bibitem[\protect\citeauthoryear{{Kovalev}, {Petrov}  \& {Plavin}}{{Kovalev} et~al.}{2017}]{2017A&A...598L...1K}
{Kovalev} Y.~Y.,  {Petrov} L.,   {Plavin} A.~V.,  2017, \mn@doi [\aap] {10.1051/0004-6361/201630031}, \href {https://ui.adsabs.harvard.edu/abs/2017A&A...598L...1K} {598, L1}

\bibitem[\protect\citeauthoryear{{Kovalev}, {Zobnina}, {Plavin}  \& {Blinov}}{{Kovalev} et~al.}{2020}]{2020MNRAS.493L..54K}
{Kovalev} Y.~Y.,  {Zobnina} D.~I.,  {Plavin} A.~V.,   {Blinov} D.,  2020, \mn@doi [\mnras] {10.1093/mnrasl/slaa008}, \href {https://ui.adsabs.harvard.edu/abs/2020MNRAS.493L..54K} {493, L54}

\bibitem[\protect\citeauthoryear{{Kr{\'a}sn{\'a}} et~al.,}{{Kr{\'a}sn{\'a}} et~al.}{2023}]{r:kra23}
{Kr{\'a}sn{\'a}} H.,  et~al., 2023, \mn@doi [\aap] {10.1051/0004-6361/202245434}, \href {https://ui.adsabs.harvard.edu/abs/2023A&A...679A..53K} {679, A53}

\bibitem[\protect\citeauthoryear{{Kravchenko}, {Pashchenko}, {Homan}, {Kovalev}, {Lister}, {Pushkarev}, {Ros}  \& {Savolainen}}{{Kravchenko} et~al.}{2025}]{2025MNRAS.538.2008K}
{Kravchenko} E.~V.,  {Pashchenko} I.~N.,  {Homan} D.~C.,  {Kovalev} Y.~Y.,  {Lister} M.~L.,  {Pushkarev} A.~B.,  {Ros} E.,   {Savolainen} T.,  2025, \mn@doi [\mnras] {10.1093/mnras/staf343}, \href {https://ui.adsabs.harvard.edu/abs/2025MNRAS.538.2008K} {538, 2008}

\bibitem[\protect\citeauthoryear{{Krezinger}, {Frey}, {An}, {Jaiswal}  \& {Zhang}}{{Krezinger} et~al.}{2020}]{2020MNRAS.496.1811K}
{Krezinger} M.,  {Frey} S.,  {An} T.,  {Jaiswal} S.,   {Zhang} Y.,  2020, \mn@doi [\mnras] {10.1093/mnras/staa1669}, \href {https://ui.adsabs.harvard.edu/abs/2020MNRAS.496.1811K} {496, 1811}

\bibitem[\protect\citeauthoryear{{Lambert}, {Liu}, {Arias}, {Barache}, {Souchay}, {Taris}, {Liu}  \& {Zhu}}{{Lambert} et~al.}{2021}]{2021A&A...651A..64L}
{Lambert} S.,  {Liu} N.,  {Arias} E.~F.,  {Barache} C.,  {Souchay} J.,  {Taris} F.,  {Liu} J.~C.,   {Zhu} Z.,  2021, \mn@doi [\aap] {10.1051/0004-6361/202140652}, \href {https://ui.adsabs.harvard.edu/abs/2021A&A...651A..64L} {651, A64}

\bibitem[\protect\citeauthoryear{{Lambert}, {Sol}  \& {Pierron}}{{Lambert} et~al.}{2024}]{2024A&A...684A.202L}
{Lambert} S.,  {Sol} H.,   {Pierron} A.,  2024, \mn@doi [\aap] {10.1051/0004-6361/202347210}, \href {https://ui.adsabs.harvard.edu/abs/2024A&A...684A.202L} {684, A202}

\bibitem[\protect\citeauthoryear{{Lee}}{{Lee}}{2014}]{2014JKAS...47..303L}
{Lee} S.-S.,  2014, \mn@doi [Journal of Korean Astronomical Society] {10.5303/JKAS.2014.47.6.303}, \href {https://ui.adsabs.harvard.edu/abs/2014JKAS...47..303L} {47, 303}

\bibitem[\protect\citeauthoryear{{Liodakis}, {Hovatta}, {Huppenkothen}, {Kiehlmann}, {Max-Moerbeck}  \& {Readhead}}{{Liodakis} et~al.}{2018}]{2018ApJ...866..137L}
{Liodakis} I.,  {Hovatta} T.,  {Huppenkothen} D.,  {Kiehlmann} S.,  {Max-Moerbeck} W.,   {Readhead} A. C.~S.,  2018, \mn@doi [\apj] {10.3847/1538-4357/aae2b7}, \href {https://ui.adsabs.harvard.edu/abs/2018ApJ...866..137L} {866, 137}

\bibitem[\protect\citeauthoryear{{Lisakov}, {Kovalev}, {Savolainen}, {Hovatta}  \& {Kutkin}}{{Lisakov} et~al.}{2017}]{2017MNRAS.468.4478L}
{Lisakov} M.~M.,  {Kovalev} Y.~Y.,  {Savolainen} T.,  {Hovatta} T.,   {Kutkin} A.~M.,  2017, \mn@doi [\mnras] {10.1093/mnras/stx710}, \href {https://ui.adsabs.harvard.edu/abs/2017MNRAS.468.4478L} {468, 4478}

\bibitem[\protect\citeauthoryear{{Lister}, {Aller}, {Aller}, {Hodge}, {Homan}, {Kovalev}, {Pushkarev}  \& {Savolainen}}{{Lister} et~al.}{2018}]{2018ApJS..234...12L}
{Lister} M.~L.,  {Aller} M.~F.,  {Aller} H.~D.,  {Hodge} M.~A.,  {Homan} D.~C.,  {Kovalev} Y.~Y.,  {Pushkarev} A.~B.,   {Savolainen} T.,  2018, \mn@doi [\apjs] {10.3847/1538-4365/aa9c44}, \href {https://ui.adsabs.harvard.edu/abs/2018ApJS..234...12L} {234, 12}

\bibitem[\protect\citeauthoryear{{Lister}, {Homan}, {Kellermann}, {Kovalev}, {Pushkarev}, {Ros}  \& {Savolainen}}{{Lister} et~al.}{2021}]{2021ApJ...923...30L}
{Lister} M.~L.,  {Homan} D.~C.,  {Kellermann} K.~I.,  {Kovalev} Y.~Y.,  {Pushkarev} A.~B.,  {Ros} E.,   {Savolainen} T.,  2021, \mn@doi [\apj] {10.3847/1538-4357/ac230f}, \href {https://ui.adsabs.harvard.edu/abs/2021ApJ...923...30L} {923, 30}

\bibitem[\protect\citeauthoryear{{Makarov}, {Frouard}, {Berghea}, {Rest}, {Chambers}, {Kaiser}, {Kudritzki}  \& {Magnier}}{{Makarov} et~al.}{2017}]{2017ApJ...835L..30M}
{Makarov} V.~V.,  {Frouard} J.,  {Berghea} C.~T.,  {Rest} A.,  {Chambers} K.~C.,  {Kaiser} N.,  {Kudritzki} R.-P.,   {Magnier} E.~A.,  2017, \mn@doi [\apjl] {10.3847/2041-8213/835/2/L30}, \href {https://ui.adsabs.harvard.edu/abs/2017ApJ...835L..30M} {835, L30}

\bibitem[\protect\citeauthoryear{{Makarov}, {Berghea}, {Frouard}, {Fey}  \& {Schmitt}}{{Makarov} et~al.}{2019}]{2019ApJ...873..132M}
{Makarov} V.~V.,  {Berghea} C.~T.,  {Frouard} J.,  {Fey} A.,   {Schmitt} H.~R.,  2019, \mn@doi [\apj] {10.3847/1538-4357/aafa1c}, \href {https://ui.adsabs.harvard.edu/abs/2019ApJ...873..132M} {873, 132}

\bibitem[\protect\citeauthoryear{{Marscher} \& {Gear}}{{Marscher} \& {Gear}}{1985}]{1985ApJ...298..114M}
{Marscher} A.~P.,  {Gear} W.~K.,  1985, \mn@doi [\apj] {10.1086/163592}, \href {https://ui.adsabs.harvard.edu/abs/1985ApJ...298..114M} {298, 114}

\bibitem[\protect\citeauthoryear{{Mignard} et~al.,}{{Mignard} et~al.}{2016}]{2016A&A...595A...5M}
{Mignard} F.,  et~al., 2016, \mn@doi [\aap] {10.1051/0004-6361/201629534}, \href {https://ui.adsabs.harvard.edu/abs/2016A&A...595A...5M} {595, A5}

\bibitem[\protect\citeauthoryear{{Mingaliev}, {Sotnikova}, {Larionov}  \& {Erkenov}}{{Mingaliev} et~al.}{2011}]{2011ARep...55..187M}
{Mingaliev} M.~G.,  {Sotnikova} Y.~V.,  {Larionov} M.~G.,   {Erkenov} A.~K.,  2011, \mn@doi [Astronomy Reports] {10.1134/S1063772911010045}, \href {https://ui.adsabs.harvard.edu/abs/2011ARep...55..187M} {55, 187}

\bibitem[\protect\citeauthoryear{{Mingaliev}, {Sotnikova}, {Mufakharov}, {Erkenov}  \& {Udovitskiy}}{{Mingaliev} et~al.}{2013}]{2013AstBu..68..262M}
{Mingaliev} M.~G.,  {Sotnikova} Y.~V.,  {Mufakharov} T.~V.,  {Erkenov} A.~K.,   {Udovitskiy} R.~Y.,  2013, \mn@doi [Astrophysical Bulletin] {10.1134/S1990341313030036}, \href {https://ui.adsabs.harvard.edu/abs/2013AstBu..68..262M} {68, 262}

\bibitem[\protect\citeauthoryear{{Murphy}, {Kacprzak}, {Savorgnan}  \& {Carswell}}{{Murphy} et~al.}{2019}]{2019MNRAS.482.3458M}
{Murphy} M.~T.,  {Kacprzak} G.~G.,  {Savorgnan} G. A.~D.,   {Carswell} R.~F.,  2019, \mn@doi [\mnras] {10.1093/mnras/sty2834}, \href {https://ui.adsabs.harvard.edu/abs/2019MNRAS.482.3458M} {482, 3458}

\bibitem[\protect\citeauthoryear{{Nan Ren-Dong}, {Schilizzi}, {van Breugel}, {Fanti}, {Fanti}, {Muxlow}  \& {Spencer}}{{Nan Ren-Dong} et~al.}{1991}]{1991A&A...245..449N}
{Nan Ren-Dong} {Schilizzi} R.~T.,  {van Breugel} W.~J.~M.,  {Fanti} C.,  {Fanti} R.,  {Muxlow} T.~W.~B.,   {Spencer} R.~E.,  1991, \aap, \href {https://ui.adsabs.harvard.edu/abs/1991A&A...245..449N} {245, 449}

\bibitem[\protect\citeauthoryear{{Nokhrina} \& {Pushkarev}}{{Nokhrina} \& {Pushkarev}}{2024}]{2024MNRAS.528.2523N}
{Nokhrina} E.~E.,  {Pushkarev} A.~B.,  2024, \mn@doi [\mnras] {10.1093/mnras/stae179}, \href {https://ui.adsabs.harvard.edu/abs/2024MNRAS.528.2523N} {528, 2523}

\bibitem[\protect\citeauthoryear{{Osetrova}, {Titov}  \& {Melnikov}}{{Osetrova} et~al.}{2024}]{2024AstL...50..657O}
{Osetrova} A.~A.,  {Titov} O.~A.,   {Melnikov} A.~E.,  2024, \mn@doi [Astronomy Letters] {10.1134/S1063773725700045}, \href {https://ui.adsabs.harvard.edu/abs/2024AstL...50..657O} {50, 657}

\bibitem[\protect\citeauthoryear{{Pashchenko}, {Kravchenko}, {Nokhrina}  \& {Nikonov}}{{Pashchenko} et~al.}{2023}]{2023MNRAS.523.1247P}
{Pashchenko} I.~N.,  {Kravchenko} E.~V.,  {Nokhrina} E.~E.,   {Nikonov} A.~S.,  2023, \mn@doi [\mnras] {10.1093/mnras/stad1527}, \href {https://ui.adsabs.harvard.edu/abs/2023MNRAS.523.1247P} {523, 1247}

\bibitem[\protect\citeauthoryear{{Petrov}}{{Petrov}}{2013}]{2013AJ....146....5P}
{Petrov} L.,  2013, \mn@doi [\aj] {10.1088/0004-6256/146/1/5}, \href {https://ui.adsabs.harvard.edu/abs/2013AJ....146....5P} {146, 5}

\bibitem[\protect\citeauthoryear{{Petrov}}{{Petrov}}{2021}]{r:wfcs}
{Petrov} L.,  2021, \mn@doi [\aj] {10.3847/1538-3881/abc4e1}, \href {https://ui.adsabs.harvard.edu/abs/2021AJ....161...14P} {161, 14}

\bibitem[\protect\citeauthoryear{{Petrov}}{{Petrov}}{2024}]{2024AJ....168...76P}
{Petrov} L.,  2024, \mn@doi [\aj] {10.3847/1538-3881/ad4a6b}, \href {https://ui.adsabs.harvard.edu/abs/2024AJ....168...76P} {168, 76}

\bibitem[\protect\citeauthoryear{{Petrov} \& {Kovalev}}{{Petrov} \& {Kovalev}}{2017a}]{2017MNRAS.467L..71P}
{Petrov} L.,  {Kovalev} Y.~Y.,  2017a, \mn@doi [\mnras] {10.1093/mnrasl/slx001}, \href {https://ui.adsabs.harvard.edu/abs/2017MNRAS.467L..71P} {467, L71}

\bibitem[\protect\citeauthoryear{{Petrov} \& {Kovalev}}{{Petrov} \& {Kovalev}}{2017b}]{2017MNRAS.471.3775P}
{Petrov} L.,  {Kovalev} Y.~Y.,  2017b, \mn@doi [\mnras] {10.1093/mnras/stx1747}, \href {https://ui.adsabs.harvard.edu/abs/2017MNRAS.471.3775P} {471, 3775}

\bibitem[\protect\citeauthoryear{{Petrov} \& {Kovalev}}{{Petrov} \& {Kovalev}}{2025}]{RFC}
{Petrov} L.~Y.,  {Kovalev} Y.~Y.,  2025, \mn@doi [\apjs] {10.3847/1538-4365/ad8c36}, \href {https://ui.adsabs.harvard.edu/abs/2025ApJS..276...38P} {276, 38}

\bibitem[\protect\citeauthoryear{{Petrov} \& {Taylor}}{{Petrov} \& {Taylor}}{2011}]{r:astro_vips}
{Petrov} L.,  {Taylor} G.~B.,  2011, \mn@doi [\aj] {10.1088/0004-6256/142/3/89}, \href {http://adsabs.harvard.edu/abs/2011AJ....142...89P} {142, 89}

\bibitem[\protect\citeauthoryear{{Petrov}, {Kovalev}, {Fomalont}  \& {Gordon}}{{Petrov} et~al.}{2011}]{2011AJ....142...35P}
{Petrov} L.,  {Kovalev} Y.~Y.,  {Fomalont} E.~B.,   {Gordon} D.,  2011, \mn@doi [\aj] {10.1088/0004-6256/142/2/35}, \href {https://ui.adsabs.harvard.edu/abs/2011AJ....142...35P} {142, 35}

\bibitem[\protect\citeauthoryear{{Petrov}, {Kovalev}  \& {Plavin}}{{Petrov} et~al.}{2019}]{2019MNRAS.482.3023P}
{Petrov} L.,  {Kovalev} Y.~Y.,   {Plavin} A.~V.,  2019, \mn@doi [\mnras] {10.1093/mnras/sty2807}, \href {https://ui.adsabs.harvard.edu/abs/2019MNRAS.482.3023P} {482, 3023}

\bibitem[\protect\citeauthoryear{{Plavin}, {Kovalev}  \& {Petrov}}{{Plavin} et~al.}{2019}]{2019ApJ...871..143P}
{Plavin} A.~V.,  {Kovalev} Y.~Y.,   {Petrov} L.~Y.,  2019, \mn@doi [\apj] {10.3847/1538-4357/aaf650}, \href {https://ui.adsabs.harvard.edu/abs/2019ApJ...871..143P} {871, 143}

\bibitem[\protect\citeauthoryear{{Plavin}, {Kovalev}  \& {Pushkarev}}{{Plavin} et~al.}{2022}]{2022ApJS..260....4P}
{Plavin} A.~V.,  {Kovalev} Y.~Y.,   {Pushkarev} A.~B.,  2022, \mn@doi [\apjs] {10.3847/1538-4365/ac6352}, \href {https://ui.adsabs.harvard.edu/abs/2022ApJS..260....4P} {260, 4}

\bibitem[\protect\citeauthoryear{{Popkov}, {Kovalev}, {Petrov}  \& {Kovalev}}{{Popkov} et~al.}{2021}]{2021AJ....161...88P}
{Popkov} A.~V.,  {Kovalev} Y.~Y.,  {Petrov} L.~Y.,   {Kovalev} Y.~A.,  2021, \mn@doi [\aj] {10.3847/1538-3881/abd18c}, \href {https://ui.adsabs.harvard.edu/abs/2021AJ....161...88P} {161, 88}

\bibitem[\protect\citeauthoryear{{Porcas}}{{Porcas}}{2009}]{2009A&A...505L...1P}
{Porcas} R.~W.,  2009, \mn@doi [\aap] {10.1051/0004-6361/200912846}, \href {https://ui.adsabs.harvard.edu/abs/2009A&A...505L...1P} {505, L1}

\bibitem[\protect\citeauthoryear{{Pushkarev} \& {Kovalev}}{{Pushkarev} \& {Kovalev}}{2012}]{2012A&A...544A..34P}
{Pushkarev} A.~B.,  {Kovalev} Y.~Y.,  2012, \mn@doi [\aap] {10.1051/0004-6361/201219352}, \href {https://ui.adsabs.harvard.edu/abs/2012A&A...544A..34P} {544, A34}

\bibitem[\protect\citeauthoryear{{Pushkarev} \& {Kovalev}}{{Pushkarev} \& {Kovalev}}{2015}]{2015MNRAS.452.4274P}
{Pushkarev} A.~B.,  {Kovalev} Y.~Y.,  2015, \mn@doi [\mnras] {10.1093/mnras/stv1539}, \href {https://ui.adsabs.harvard.edu/abs/2015MNRAS.452.4274P} {452, 4274}

\bibitem[\protect\citeauthoryear{{Pushkarev}, {Hovatta}, {Kovalev}, {Lister}, {Lobanov}, {Savolainen}  \& {Zensus}}{{Pushkarev} et~al.}{2012}]{2012AA...545A.113P}
{Pushkarev} A.~B.,  {Hovatta} T.,  {Kovalev} Y.~Y.,  {Lister} M.~L.,  {Lobanov} A.~P.,  {Savolainen} T.,   {Zensus} J.~A.,  2012, \mn@doi [\aap] {10.1051/0004-6361/201219173}, \href {https://ui.adsabs.harvard.edu/abs/2012A&A...545A.113P} {545, A113}

\bibitem[\protect\citeauthoryear{{Readhead}}{{Readhead}}{1994}]{1994ApJ...426...51R}
{Readhead} A. C.~S.,  1994, \mn@doi [\apj] {10.1086/174038}, \href {https://ui.adsabs.harvard.edu/abs/1994ApJ...426...51R} {426, 51}

\bibitem[\protect\citeauthoryear{{Richards} et~al.,}{{Richards} et~al.}{2009}]{2009ApJS..180...67R}
{Richards} G.~T.,  et~al., 2009, \mn@doi [\apjs] {10.1088/0067-0049/180/1/67}, \href {https://ui.adsabs.harvard.edu/abs/2009ApJS..180...67R} {180, 67}

\bibitem[\protect\citeauthoryear{{R{\"o}der} et~al.,}{{R{\"o}der} et~al.}{2025}]{2025A&A...695A.233R}
{R{\"o}der} J.,  et~al., 2025, \mn@doi [\aap] {10.1051/0004-6361/202452600}, \href {https://ui.adsabs.harvard.edu/abs/2025A&A...695A.233R} {695, A233}

\bibitem[\protect\citeauthoryear{{Rodriguez}, {Taylor}, {Zavala}, {Peck}, {Pollack}  \& {Romani}}{{Rodriguez} et~al.}{2006}]{2006ApJ...646...49R}
{Rodriguez} C.,  {Taylor} G.~B.,  {Zavala} R.~T.,  {Peck} A.~B.,  {Pollack} L.~K.,   {Romani} R.~W.,  2006, \mn@doi [\apj] {10.1086/504825}, \href {https://ui.adsabs.harvard.edu/abs/2006ApJ...646...49R} {646, 49}

\bibitem[\protect\citeauthoryear{{Schinzel}, {Lobanov}, {Taylor}, {Jorstad}, {Marscher}  \& {Zensus}}{{Schinzel} et~al.}{2012}]{2012A&A...537A..70S}
{Schinzel} F.~K.,  {Lobanov} A.~P.,  {Taylor} G.~B.,  {Jorstad} S.~G.,  {Marscher} A.~P.,   {Zensus} J.~A.,  2012, \mn@doi [\aap] {10.1051/0004-6361/201117705}, \href {https://ui.adsabs.harvard.edu/abs/2012A&A...537A..70S} {537, A70}

\bibitem[\protect\citeauthoryear{{Secrest}}{{Secrest}}{2022}]{2022ApJ...939L..32S}
{Secrest} N.~J.,  2022, \mn@doi [\apjl] {10.3847/2041-8213/ac8d5d}, \href {https://ui.adsabs.harvard.edu/abs/2022ApJ...939L..32S} {939, L32}

\bibitem[\protect\citeauthoryear{{Shepherd}}{{Shepherd}}{1997}]{1997ASPC..125...77S}
{Shepherd} M.~C.,  1997, in {Hunt} G.,  {Payne} H.,  eds,  Astronomical Society of the Pacific Conference Series Vol. 125, Astronomical Data Analysis Software and Systems VI. p.~77

\bibitem[\protect\citeauthoryear{{Shepherd}, {Pearson}  \& {Taylor}}{{Shepherd} et~al.}{1994}]{1994BAAS...26..987S}
{Shepherd} M.~C.,  {Pearson} T.~J.,   {Taylor} G.~B.,  1994, Bulletin of the American Astronomical Society, \href {https://ui.adsabs.harvard.edu/abs/1994BAAS...26..987S} {26, 987}

\bibitem[\protect\citeauthoryear{{Sokolovsky}, {Kovalev}, {Kovalev}, {Nizhelskiy}  \& {Zhekanis}}{{Sokolovsky} et~al.}{2009}]{2009AN....330..199S}
{Sokolovsky} K.~V.,  {Kovalev} Y.~Y.,  {Kovalev} Y.~A.,  {Nizhelskiy} N.~A.,   {Zhekanis} G.~V.,  2009, \mn@doi [Astronomische Nachrichten] {10.1002/asna.200811155}, \href {https://ui.adsabs.harvard.edu/abs/2009AN....330..199S} {330, 199}

\bibitem[\protect\citeauthoryear{{Sokolovsky}, {Kovalev}, {Pushkarev}, {Mimica}  \& {Perucho}}{{Sokolovsky} et~al.}{2011}]{2011A&A...535A..24S}
{Sokolovsky} K.~V.,  {Kovalev} Y.~Y.,  {Pushkarev} A.~B.,  {Mimica} P.,   {Perucho} M.,  2011, \mn@doi [\aap] {10.1051/0004-6361/201015772}, \href {https://ui.adsabs.harvard.edu/abs/2011A&A...535A..24S} {535, A24}

\bibitem[\protect\citeauthoryear{{Sotnikova}, {Mufakharov}, {Majorova}, {Mingaliev}, {Udovitskii}, {Bursov}  \& {Semenova}}{{Sotnikova} et~al.}{2019}]{2019AstBu..74..348S}
{Sotnikova} Y.~V.,  {Mufakharov} T.~V.,  {Majorova} E.~K.,  {Mingaliev} M.~G.,  {Udovitskii} R.~Y.,  {Bursov} N.~N.,   {Semenova} T.~A.,  2019, \mn@doi [Astrophysical Bulletin] {10.1134/S1990341319040023}, \href {https://ui.adsabs.harvard.edu/abs/2019AstBu..74..348S} {74, 348}

\bibitem[\protect\citeauthoryear{{Strauss}, {Huchra}, {Davis}, {Yahil}, {Fisher}  \& {Tonry}}{{Strauss} et~al.}{1992}]{1992ApJS...83...29S}
{Strauss} M.~A.,  {Huchra} J.~P.,  {Davis} M.,  {Yahil} A.,  {Fisher} K.~B.,   {Tonry} J.,  1992, \mn@doi [\apjs] {10.1086/191730}, \href {https://ui.adsabs.harvard.edu/abs/1992ApJS...83...29S} {83, 29}

\bibitem[\protect\citeauthoryear{{Sulentic}, {Stirpe}, {Marziani}, {Zamanov}, {Calvani}  \& {Braito}}{{Sulentic} et~al.}{2004}]{2004A&A...423..121S}
{Sulentic} J.~W.,  {Stirpe} G.~M.,  {Marziani} P.,  {Zamanov} R.,  {Calvani} M.,   {Braito} V.,  2004, \mn@doi [\aap] {10.1051/0004-6361:20035912}, \href {https://ui.adsabs.harvard.edu/abs/2004A&A...423..121S} {423, 121}

\bibitem[\protect\citeauthoryear{{Titov}, {Stanford}, {Johnston}, {Pursimo}, {Hunstead}, {Jauncey}, {Maslennikov}  \& {Boldycheva}}{{Titov} et~al.}{2013}]{2013AJ....146...10T}
{Titov} O.,  {Stanford} L.~M.,  {Johnston} H.~M.,  {Pursimo} T.,  {Hunstead} R.~W.,  {Jauncey} D.~L.,  {Maslennikov} K.,   {Boldycheva} A.,  2013, \mn@doi [\aj] {10.1088/0004-6256/146/1/10}, \href {https://ui.adsabs.harvard.edu/abs/2013AJ....146...10T} {146, 10}

\bibitem[\protect\citeauthoryear{{Titov} et~al.,}{{Titov} et~al.}{2022}]{2022MNRAS.512..874T}
{Titov} O.,  et~al., 2022, \mn@doi [\mnras] {10.1093/mnras/stac038}, \href {https://ui.adsabs.harvard.edu/abs/2022MNRAS.512..874T} {512, 874}

\bibitem[\protect\citeauthoryear{{Verkhodanov}, {Trushkin}, {Andernach}  \& {Chernenkov}}{{Verkhodanov} et~al.}{2005}]{2005BSAO...58..118V}
{Verkhodanov} O.~V.,  {Trushkin} S.~A.,  {Andernach} H.,   {Chernenkov} V.~N.,  2005, \mn@doi [Bulletin of the Special Astrophysics Observatory] {10.48550/arXiv.0705.2959}, \href {https://ui.adsabs.harvard.edu/abs/2005BSAO...58..118V} {58, 118}

\bibitem[\protect\citeauthoryear{{Wilkinson}, {Polatidis}, {Readhead}, {Xu}  \& {Pearson}}{{Wilkinson} et~al.}{1994}]{1994ApJ...432L..87W}
{Wilkinson} P.~N.,  {Polatidis} A.~G.,  {Readhead} A.~C.~S.,  {Xu} W.,   {Pearson} T.~J.,  1994, \mn@doi [\apjl] {10.1086/187518}, \href {https://ui.adsabs.harvard.edu/abs/1994ApJ...432L..87W} {432, L87}

\bibitem[\protect\citeauthoryear{{Xu} \& {Charlot}}{{Xu} \& {Charlot}}{2025}]{2025AJ....169..173X}
{Xu} M.~H.,  {Charlot} P.,  2025, \mn@doi [\aj] {10.3847/1538-3881/adb133}, \href {https://ui.adsabs.harvard.edu/abs/2025AJ....169..173X} {169, 173}

\bibitem[\protect\citeauthoryear{{Xu} \& {Han}}{{Xu} \& {Han}}{2014}]{2014MNRAS.442.3329X}
{Xu} J.,  {Han} J.~L.,  2014, \mn@doi [\mnras] {10.1093/mnras/stu1018}, \href {https://ui.adsabs.harvard.edu/abs/2014MNRAS.442.3329X} {442, 3329}

\bibitem[\protect\citeauthoryear{{Xu}, {Lunz}, {Anderson}, {Savolainen}, {Zubko}  \& {Schuh}}{{Xu} et~al.}{2021}]{2021A&A...647A.189X}
{Xu} M.~H.,  {Lunz} S.,  {Anderson} J.~M.,  {Savolainen} T.,  {Zubko} N.,   {Schuh} H.,  2021, \mn@doi [\aap] {10.1051/0004-6361/202040168}, \href {https://ui.adsabs.harvard.edu/abs/2021A&A...647A.189X} {647, A189}

\bibitem[\protect\citeauthoryear{{Xu}, {Savolainen}, {Anderson}, {Kareinen}, {Zubko}, {Lunz}  \& {Schuh}}{{Xu} et~al.}{2022}]{2022A&A...663A..83X}
{Xu} M.~H.,  {Savolainen} T.,  {Anderson} J.~M.,  {Kareinen} N.,  {Zubko} N.,  {Lunz} S.,   {Schuh} H.,  2022, \mn@doi [\aap] {10.1051/0004-6361/202140840}, \href {https://ui.adsabs.harvard.edu/abs/2022A&A...663A..83X} {663, A83}

\bibitem[\protect\citeauthoryear{{Xu} et~al.,}{{Xu} et~al.}{2024}]{2024AJ....168..219X}
{Xu} S.,  et~al., 2024, \mn@doi [\aj] {10.3847/1538-3881/ad7af0}, \href {https://ui.adsabs.harvard.edu/abs/2024AJ....168..219X} {168, 219}

\bibitem[\protect\citeauthoryear{{Yao} et~al.,}{{Yao} et~al.}{2019}]{2019ApJS..240....6Y}
{Yao} S.,  et~al., 2019, \mn@doi [\apjs] {10.3847/1538-4365/aaef88}, \href {https://ui.adsabs.harvard.edu/abs/2019ApJS..240....6Y} {240, 6}

\bibitem[\protect\citeauthoryear{{Zamaninasab}, {Savolainen}, {Clausen-Brown}, {Hovatta}, {Lister}, {Krichbaum}, {Kovalev}  \& {Pushkarev}}{{Zamaninasab} et~al.}{2013}]{2013MNRAS.436.3341Z}
{Zamaninasab} M.,  {Savolainen} T.,  {Clausen-Brown} E.,  {Hovatta} T.,  {Lister} M.~L.,  {Krichbaum} T.~P.,  {Kovalev} Y.~Y.,   {Pushkarev} A.~B.,  2013, \mn@doi [\mnras] {10.1093/mnras/stt1816}, \href {https://ui.adsabs.harvard.edu/abs/2013MNRAS.436.3341Z} {436, 3341}

\makeatother
\end{thebibliography}


\bsp	
\label{lastpage}
\end{document}